\documentclass[a4paper,11pt]{article}
\pdfoutput=1 

\usepackage{jheppub} 

\usepackage[T1]{fontenc} 

\usepackage[utf8]{inputenc}
\usepackage{graphicx} 
 \graphicspath{{./figures/}}
\usepackage{dcolumn} 
\usepackage{amsmath,amsfonts} 
\allowdisplaybreaks

\usepackage{mathabx}
\usepackage{physics}
\usepackage{xcolor}
\usepackage{slashed}
\usepackage{bm}
\usepackage{cancel}
\usepackage[caption=false]{subfig}
\usepackage{wrapfig}
\usepackage[dvipsnames,svgnames]{xcolor}
\usepackage{braket}
\usepackage{multirow}
\usepackage{makecell}
\usepackage{graphicx}
\usepackage{hhline}
\usepackage[shortlabels]{enumitem}

\newcommand{\beq}{\begin{equation}}
\newcommand{\eeq}{\end{equation}}

\usepackage{hyperref} 

\usepackage[capitalise]{cleveref}

\hypersetup{%
  colorlinks=true, linktocpage=true, pdfstartpage=1, pdfstartview=FitV,%
  breaklinks=true, pdfpagemode=UseNone, pageanchor=true, pdfpagemode=UseOutlines,%
  plainpages=false, bookmarksnumbered, bookmarksopen=true, bookmarksopenlevel=1,%
  hypertexnames=true, pdfhighlight=/O,
  urlcolor=Cerulean, linkcolor=Cerulean, citecolor=Cerulean,%
}

\def\Tp{T^\prime}
\usepackage{framed}
\colorlet{shadecolor}{blue!20}
\usepackage[normalem]{ulem}

\makeatletter
\gdef\@fpheader{}
\makeatother

\begin{document}

\title{Probing the Phenomenology of Dark Matter from Decoupled Freeze-Out}
\author[a]{Geneviève Bélanger,}
\emailAdd{belanger@lapth.cnrs.fr}
\affiliation[a]{LAPTh, 
            USMB, CNRS, F-74940 Annecy, France}
\author[b]{Aoife Bharucha,}
\emailAdd{aoife.bharucha@cpt.univ-mrs.fr}
\affiliation[b]{Aix Marseille Univ,
            Université de Toulon,
            CNRS, CPT, Marseille, France}
\author[c]{Sreemanti Chakraborti,}
\emailAdd{sreemanti.chakraborti@durham.ac.uk}
\affiliation[c]{Institute for Particle Physics Phenomenology,
            Department of Physics,
            Durham University,
            Durham DH1 3LE, United Kingdom}

\author[d]{Rashidul Islam}
\emailAdd{islam.rashid@gmail.com}
\affiliation[d]{Mathabhanga College,
            Cooch Behar Panchanan Barma University,
            Cooch Behar 736146, India}
\author[f]{and Sophie Mutzel}
\emailAdd{sophie.mutzel@minesparis.psl.eu}
\affiliation[f]{Laboratoire de Physique de l’École Normale Supérieure, Mines Paris - PSL, Inria, CNRS, ENS-
PSL, Sorbonne Université, PSL Research University, Paris, France}

\abstract{We consider a model of dark matter where the mediator corresponds to a superposition of a scalar and pseudoscalar, and the scenario where, after reheating, the number densities of the dark sector particles, i.e.~the dark matter and the mediators, are negligible.
If the coupling of the mediators to the Standard Model is feeble, but the coupling to the dark matter is large enough, the dark sector may reach equilibrium at a temperature distinct from that of the thermal bath.
The relic density is then said to be obtained via decoupled freeze out (DFO). We focus on the $s$-wave annihilation scenario, which particularly benefits from the DFO mechanism by evading standard CMB limits while still yielding indirect detection signals.
We calculate the relic density by solving a set of four coupled Boltzmann equations for the number densities of the dark sector particles and the energy transfer from the light to dark sector.
We finally perform a thorough analysis of experimental bounds on this scenario, namely from indirect detection and the CMB, as well as from BBN, and find that, while there are considerable constraints on the parameter space where the correct relic density is obtained, a viable region remains to be explored.}

\preprint{IPPP/25/79}

\maketitle

%

\section{Introduction}
Thermal freeze-out has long been considered the most attractive mechanism for producing dark matter (DM) in the early Universe.  However, the stringent constraints on  weakly interacting massive particles  (WIMP)~\cite{Cirelli:2024ssz} as dark matter derived from  a wide range of experimental searches have rekindled  the interest in alternative DM production mechanisms  that cover a much wider range of masses and coupling strengths.
Of special interest are the cases that allow light DM (below the electroweak scale) and/or DM that is very weakly (even feebly coupled)~\cite{McDonald:2001vt,Hall:2009bx,Bernal:2017kxu}.
 
One determining factor in dark matter formation is whether  DM  reaches thermal equilibrium in the early universe, as is the case for WIMPs. If it does, apart from standard freeze-out (FO), where DM annihilation into Standard Model (SM) particles determines the DM relic density, other thermal processes include secluded FO~\cite{Pospelov:2007mp,Borah:2025fkd}  where DM annihilates into dark states, forbidden FO~\cite{DAgnolo:2015ujb,DAgnolo:2020mpt} where the final states are heavier than DM, assisted FO~\cite{Belanger:2011ww} or multi-component FO~\cite{Zurek:2008qg,Liu:2011aa,Belanger:2012vp,Esch:2014jpa,Arcadi:2016kmk,Chakraborti:2018lso, Chakraborti:2018aae, Belanger:2021lwd,Belanger:2022esk} where FO is affected by the interactions between at least two dark sectors as well as by the interactions of the dark sectors with the SM. When the DM is so weakly coupled that it never reaches equilibrium with the mediator or the SM bath, the  freeze-in mechanism~\cite{McDonald:2001vt,Hall:2009bx} comes into play, and dark matter is slowly produced from the decay or annihilation of particles in the  thermal bath. Other non-thermal mechanisms include production from out-of-equilibrium decays of frozen-out WIMPs (the superWIMP)~\cite{Feng:2003uy}, from sequential freeze-in~\cite{Hambye:2019dwd, Belanger:2020npe}, as well as from ``leak-in" and glaciation scenarios~\cite{Evans:2019vxr,Fernandez:2021iti}.  Dark matter formation in scenarios where  the dark and visible sector couple only gravitationally  have also been considered in ~\cite{Cheung:2010gj}.
 
For couplings that are weaker than those of the WIMPs yet not feeble, one can encounter a loss of kinetic equilibrium. Typically this occurs when the dark matter couplings to the SM drop below roughly  $10^{-6}$ as encountered, for example, when DM annihilates near  a resonance~\cite{Binder:2017rgn,Binder:2021bmg,Binder:2022pmf, Belanger:2024bro,Belanger:2025kce}, or in the conversion driven FO (coscattering)  mechanism ~\cite{Garny:2017rxs,DAgnolo:2017dbv,Brummer:2019inq,Alguero:2022inz,Chatterjee:2025vdz}. Kinetic decoupling of WIMPs in non-standard cosmological scenarios, such
as kination or low temperature reheating
cosmology, was explored in \cite{Visinelli:2015eka}. 
In another class of scenarios where the mediator is feebly coupled to the SM but the dark sector coupling is larger, before it decouples, DM can enter into equilibrium with the mediator at
a temperature distinct from that of the thermal bath. Processes that govern DM formation  are interactions of the DM with the hidden sector. 

This mechanism ~\cite{Cheung:2010gj, Chu:2011be,Hambye:2019dwd,Hambye:2020lvy,Coy:2021ann} can be viewed as a subclass of secluded FO with the important difference that the DM temperature ($T'$) is below that of the  SM bath, $T'<T$ . 
 In the following, as in ~\cite{Bharucha:2022lty}, we will generically call the mechanism of freeze-out from a thermally decoupled dark sector, decoupled freeze-out (DFO). This
includes both the cases where the dark sector has been created from SM-DM interactions ~\cite{Feng:2008mu,Cheung:2010gj, Chu:2011be} or from SM-mediator interactions ~\cite{Hambye:2019dwd,Bharucha:2022lty}.

In this paper we study the DFO mechanism, focusing on DM at or below the electroweak scale. DM production via the DFO mechanism in the case of fermionic dark matter coupled to a pseudoscalar mediator was studied in~\cite{Bharucha:2022lty}  in a scenario of axion-like particle which have typically  suppressed couplings to standard fermions. This required a derivation of the energy transfer Boltzmann equation for the case of a single mediator in the final state. In \cite{Chu:2011be} the energy transfer Boltzmann equation was derived for the case where the dark photon is a mediator, i.e.~for dark matter pair production. In \cite{Hambye:2019dwd}, the case of the production of a single mediator from the SM was taken into account, however the collision term for this energy transfer equation was not explicitly given nor commented on in detail. For that reason the energy transfer Boltzmann equation for this case was derived explicitly in \cite{Bharucha:2022lty}.
While the mediator, feebly coupled to the SM, leads to significant Big Bang Nucleosynthesis (BBN) constraints and negligible direct detection bounds, DM annihilation in the pure pseudoscalar mediator scenario is $p$-wave.
As a result, the annihilation cross section becomes highly suppressed at low velocities. Consequently, such scenarios readily evade cosmological constraints and are difficult to probe with indirect detection experiments.
We instead consider a minimal scenario that offer potential signatures in indirect detection, that is a scenario  where DM production is an $s$-wave process and is therefore not suppressed in galaxies or in the early Universe.
We therefore consider a dark sector consisting of fermionic dark matter and two mediator particles, a scalar and a pseudoscalar. 
As mentioned earlier, in the DFO scenario the interactions among the hidden sector particles, DM, scalar, and pseudoscalar, are sufficiently strong to allow for thermalization within the hidden sector, while interactions between the hidden sector and the SM particles remain feeble, leading to a slow transfer of energy from the SM sector to the hidden sector. 
We always assume that the DM is heavier than the scalar/pseudoscalar mediators which we consider to be mass degenerate for simplicity. The main mechanism which sets the DM relic density is annihilation  into a scalar and a pseudoscalar, which is an $s$-wave process. 
In this model other possible DM generation mechanisms include secluded FO (when all particles are in thermal equilibrium), freeze-in (when DM is feebly coupled to the SM and the mediator) and sequential freeze-in (where the dark matter is produced via scatterings of out-of-equilibrium mediators).

This simple model containing both weak and feeble couplings has specific  phenomenological implications for cosmology and astrophysics. 
It is well known that the annihilation of DM particles can inject entropy into the primordial plasma, leading to anisotropies in the Cosmic Microwave Background (CMB)~\cite{Slatyer:2015jla}. Typically, the measurements of anisotropies by the \texttt{Planck} satellite~\cite{Planck:2018vyg} constrain $s$-wave thermal dark matter with masses ${\cal O}(10)~\rm{GeV}$.  We will show that the CMB constraint can be relaxed in DFO. Indeed, since thermal equilibrium in the dark sector is reached for a lower temperature than in the visible sector, freeze-out requires a smaller dark-sector coupling. Thus the  DFO  mechanism can allow scenarios of light dark matter with $s$-wave annihilation\footnote{ In some $T' < T$ scenarios, cosmological constraints have been discussed, such as $\Delta N_{\rm eff}$ in DFO-like setups~\cite{Coy:2021ann}, and CMB anisotropy bounds in scenarios~\cite{Berger:2016vxi} where a mediator is produced non-thermally from the SM bath and subsequently decays into SM particles, closely resembling a ``sequential freeze-in'' mechanism~\cite{Belanger:2020npe,Hambye:2019dwd}.}.
Moreover, late decays of long-lived mediators can impact the successful predictions of BBN, leading to a lower bound on the couplings of the mediators to SM and since this coupling controls the rate of energy transfer between the visible and dark sectors, by insisting on the correct relic density, it will  indirectly  give a mass dependent lower bound on the coupling of the mediators to DM. Conversely, indirect detection of DM as well as CMB measurements will constrain larger values of the mediator coupling to DM. 
For benchmark points where the DM is always heavier than the mediator, we determine the region of parameter space which is compatible with the relic density constraint  in the decoupled freeze-out scenario and also investigate the impact of various cosmological and astrophysical constraints, namely \texttt{Fermi-LAT}~\cite{Fermi-LAT:2015att} observations of photons from Dwarf Spheroidal galaxies, radio constraints measuring synchrotron emissions from galaxy clusters as observed by the telescope \texttt{MeerKAT}~\cite{Knowles:2021nvz,Jonas:2018Jr,Beck:2023oza} and constraints from the CMB power spectrum. 
We will show that for given masses of DM and mediators, there remain regions of couplings of the DM to the mediator where the DFO mechanism works, whereas the thermal scenario of secluded freeze-out  is ruled out~\cite{Pospelov:2007mp}.
 In this work we neglect thermal corrections. In a simpler model with only a pseudoscalar mediator, the thermal corrections were found to be at most of ${\cal O}(1)$ and did not have any qualitative  impact on the phenomenology of the model~\cite{Bharucha:2022lty}, we expect the same will hold here. Indeed since the scalar/pseudoscalar are feebly coupled to the standard model the thermal corrections to the masses are expected to be small. Furthermore, in this production mechanism the dominant effect of
thermal corrections is in the energy transfer from the SM to the dark sector via
the scattering of gluons and fermions, the substantial thermal mass of the gluon
induces changes in the temperature of the dark sector which then influences the
final relic density. This is therefore a more indirect, subleading effect in contrast
with the dark photon model where thermal effects can be important, indeed the
dark photon gets corrections through in medium mixing and thermal effects modify
the photon propagator and leads to resonance enhancement of DM production~\cite{Hambye:2019dwd}. A more thorough investigation of the parameter space of the model that also includes thermal effects is left for a future publication.

The paper is organized as follows. In Sec.~\ref{sec:model} we describe the model. In Sec.~\ref{sec:DFO} we detail the calculation of the relic density including the effect from the energy transfer between the dark and SM sectors. Sec.~\ref{sec:constraints} enumerates all the relevant constraints and we discuss the numerical results for the relic density, CMB, indirect detection, and BBN constraints in Sec.~\ref{sec:results}. In Sec.~\ref{sec:secluded} we make a comparison with the case of secluded freeze-out where all particles are in thermal equilibrium. Sec.~\ref{sec:discussion} contains further discussion and our conclusions. Some details of the benchmarks are tabulated in the appendix.

%
\section{Model}\label{sec:model}

The model consists of a Dirac fermionic DM candidate $\chi$ which interacts with the SM fermions via mediators: $\phi$ (scalar) and $a$ (axion-like pseudo-scalar, i.e., an ALP) through effective interactions
\begin{align}
    {\cal L}_\text{DS-ferm}
    &=
    - c_{\phi\chi} \frac{m_\chi}{\varLambda} \phi \widebar\chi\chi
    - c_{\phi f} \frac{m_f}{\varLambda} \phi \widebar f f
    \notag\\
    &- c_{a\chi} \frac{m_\chi}{\varLambda} a \widebar\chi i\gamma_5 \chi
    - c_{af} \frac{m_f}{\varLambda} a \widebar f i\gamma_5 f\,.
    \label{eq:LagDS-ferm}
\end{align}
Note that $\chi$ is odd under a discrete $Z_2$ symmetry while all other fields are even under it. 
A possible UV completion of such interactions has been discussed in the context of TeV-scale thermal freeze-out scenarios~\cite{Nomura:2008ru}. We restrict ourselves to the leading order effectively renormalizable fermionic portal operators, assuming the DM and mediator masses, as well as the reheating temperature, to lie sufficiently below the EFT cut-off that higher-dimensional terms can in general be consistently neglected. The explicit fermion mass factors arise after electroweak symmetry breaking, preserving the gauge invariance of the SM.
Note that $\varLambda$ in the above Lagrangian is the scale of the effective field theory, $m_\chi$ and $m_f$ are the masses of the DM $\chi$ and SM fermions $f$, respectively. For the sake of economy, we write the effective couplings above as $c_{S\chi} = \varLambda\cdot g_{S\chi}$, where $S$ can be either $\phi$ or $a$. Thus, the Lagrangian \eqref{eq:LagDS-ferm} takes the form
\begin{align}
    {\cal L}_\text{DS-ferm}
    &=
    - g_{\phi\chi}\,m_\chi\,\phi\widebar\chi\chi
    - g_{\phi f}\,m_f\,\phi\widebar{f} f
    \notag\\
    &- g_{a\chi}\,m_\chi\,a \widebar\chi i\gamma_5 \chi
    - g_{af}\,m_f\,a \widebar f i\gamma_5 f\,.
    \label{eq:LagDS-ferm2}
\end{align}
The dark sector scalars interact with SM Higgs doublet $H$ ($h$ denotes the SM Higgs boson) via the potential
\begin{align}
    V (\phi, a, h)
    &=
   \frac{\lambda_{\phi h}}{2} \phi H^\dag H+ \frac{\lambda'_{\phi h}}{2} \phi^2 H^\dag H
    + \frac{\lambda_{ah}}{2} a^2 H^\dag H
    \notag\\
    &+ \lambda_{\phi a} m_\phi \phi a^2 + \frac{\lambda_{\phi\phi}}{4} \phi^4
    + \frac{\lambda_{aa}}{4} a^4 + \frac{\lambda'_{\phi a}}{2} \phi^2 a^2\,.
    \label{eq:potScal}
\end{align}
We do not introduce Higgs–mediator interactions in this work, which can be avoided by taking the portal couplings sufficiently small, in practice we set~$\lambda_{\phi h}, \lambda'_{\phi h}\, {\rm and}\ \lambda_{ah}=0$.
This choice does not affect our conclusions, provided that the interactions of the mediator with Higgs bosons are smaller than those with the SM fermions.
We also do not consider any direct coupling between the dark sector particles with the SM gauge bosons. However, the presence of scalar-fermion coupling means that they nonetheless interact with gauge bosons via loops, giving rise to the effective interactions

\begin{align}
    {\cal L}_\text{DS-gluon}
    &=
    \bigg[\sum_q g_{\phi q} F^q_G\bigg] \frac{\alpha_S(m^2_\phi)}{4\pi} \phi \tr\!\big\{{\bm G}_{\mu\nu} {\bm G}^{\mu\nu}\big\}
    \notag\\
    &+ \bigg[\sum_q g_{aq} \widetilde{F}^q_G\bigg]\frac{\alpha_S(m^2_a)}{4\pi} a \tr\!\big\{{\bm G}_{\mu\nu} \widetilde{\bm G}^{\mu\nu}\big\}.
    \label{eq:LagDS-gluon}
\end{align}
As can be seen, the above interactions are proportional to the respective gauge coupling times a form factor $F$ solely dependent the fermion masses in the loop. Hence, we kept only the gluon terms in our study \footnote{Note, however, that these loop-induced gluon couplings mainly affect indirect detection via hadronic final states. They do not play a significant role in the early-universe processes such as energy transfer or hidden sector freeze-out, which are governed by the tree-level fermionic couplings.}. Here $F^q_G, \widetilde{F}^q_G$ are form factors coming from quark loops which is given by~\cite{Djouadi:2005gi,Djouadi:2005gj}~\footnote{Similarly, there should also be mediator-photon couplings through fermion loops. We estimated the form factors to be suppressed by nearly two orders of magnitude, therefore we neglect this interaction in our analysis.}
\begin{align}
\begin{aligned}
    F^q_G
    &=
    \tau_q \big[1 + (1 - \tau_q) f(\tau_q)\big]\,;
    \quad&
    \tau_q
    =
    4 m^2_q/m^2_\phi\,,
    \\
    \widetilde{F}^q_G
    &=
    \tau_q f(\tau_q)\,;
    &
    \tau_q
    =
    4 m^2_q/m^2_a\,,
\end{aligned}
\end{align}
where
\begin{align}
  f(\tau)
    &=
    \begin{cases}
    \phantom{- \dfrac{1}{4}} \bigg[\sin^{-1}\dfrac{1}{\sqrt{\tau}}\bigg]^2    & \text{for } \tau \geq 1\,,
    \\[1em]
    - \dfrac{1}{4} \bigg[\log\dfrac{1 + \sqrt{1 - \tau}}{1 - \sqrt{1 - \tau}} - i\pi\bigg]^2    & \text{for } \tau < 1\,.
    \end{cases}  
\end{align}
As a result, the Lagrangian of our model can be written as
\begin{align}\label{eq:Lag}
    {\cal L}
    \supset
    - V (\phi, a, h)
    + {\cal L}_\text{DS-ferm}
    + {\cal L}_\text{DS-gluon}\,.
\end{align}

In summary, the model is defined by the external mass parameters $(m_\chi, m_\phi, m_a)$ and the effective couplings $(g_{\phi\chi}, g_{a\chi}, g_{\phi f}, g_{a f})$, which fully determine the relic density and indirect detection phenomenology explored in this work. 
The remaining scalar self couplings ($\lambda_{\phi a}$, $\lambda_{\phi a}^\prime$, $\lambda_{\phi\phi}$, $\lambda_{aa}$) are taken to be very small in order to highlight the role of the Yukawa couplings in the phenomenology. 
This illustrates the case where the dark sector interactions are dominated by the dark matter-mediator interactions, other possibilities are discussed in Sec.~\ref{sec:discussion}.

%
\section{Decoupled freeze-out mechanism}
\label{sec:DFO}

In our model, the observed DM relic density can be produced through various thermal, non-thermal, and in-between phases, depending on the strength of the couplings in the  Lagrangian described by Eq.~\eqref{eq:LagDS-ferm2}. These couplings essentially control the interaction strengths within the dark sector as well as between the dark sector and the SM, leading to different DM production phases~\cite{Chu:2011be,Hambye:2019dwd,Belanger:2020npe,Bharucha:2022lty} through portals. We focus on a scenario in which the hidden sector begins with a negligible initial abundance of particles and evolves as a decoupled dark sector. In this scenario, the interactions among the hidden sector particles -- the pseudoscalar~($a$), the scalar~($\phi$), and the DM particle~($\chi$) -- are sufficiently strong to establish thermal equilibrium within the sector, characterized by a shared dark temperature $\Tp$. However, the interactions between the hidden sector and SM particles are too feeble to equilibrate the two sectors, leading to a slow, out-of-equilibrium population of the hidden sector from SM scatterings and decays. 

 This idea was introduced in the context of hidden photon models in Refs.~\cite{Chu:2011be,Hambye:2019dwd}, and the names given for the production mechanisms were either  ``reannihilation'' or secluded freeze-out from a hidden sector at a different temperature,   see also
  ~\cite{Feng:2008mu, Cheung:2010gj,Chu:2011be}.
 In the first mechanism the production of DM from the SM is important at the time of decoupling and leads to an important increase in the DM abundance. The  second mechanism was further explored 
in Ref.~\cite{Bharucha:2022lty} for the case of the ALP mediator and was called ``decoupled freeze-out (DFO)",  in this scenario the production of DM comes predominantly from the mediator rather than from the SM. 
In this work, we extend the DFO scenario with two mediators with their spins chosen such that the hidden sector annihilation is an $s$-wave process. This significantly boosts the detection prospects of such scenarios, as we will elaborate in Sec.~\ref{sec:constraints}. In the relic density evolution as well, there are some subtleties that we will discuss in this section.

Thermalization within the dark sector occurs once the interaction rates among the hidden sector particles exceed the expansion rate of the universe,
\begin{equation}
\Gamma_{HS} \equiv \ev*{ \sigma_{i j \to k l}\, v} n^{\mathrm{eq}}_i(T^\prime) \gtrsim H\,,
\label{gammaannihH}
\end{equation}
where $i,j,k,l$ denote hidden sector particles involved in the process. In principle, the equilibrium condition \eqref{gammaannihH} depends non-trivially on the magnitude of multiple couplings of the Lagrangian \eqref{eq:Lag}, including $\lambda_{\phi a}$, $\lambda_{\phi a}^\prime$, $\lambda_{\phi\phi}$, $\lambda_{aa}$, $g_{\phi\chi}$ and $g_{a\chi}$. However, as discussed in the previous section, we assume that the Yukawa couplings between the (pseudo-)scalar and the DM dominate, while self-interactions $\lambda_{\phi\phi}$, $\lambda_{aa}$ as well as the direct interaction between the scalar and pseudoscalar $\lambda_{\phi a}$, $\lambda_{\phi a}^\prime$ are negligibly small.  For simplicity, we take mass degenerate $a$ and $\phi$ and also assume that the Yukawa couplings satisfy $g_{\phi\chi}\approx g_{a\chi}$ and $g_{af}\approx g_{\phi f}$, i.e.~the interaction strengths of the scalar and the pseudoscalar are comparable. Under these assumptions, and provided that $g_{\phi\chi}\approx g_{a\chi}$ are large enough to satisfy Eq.~\eqref{gammaannihH}, all hidden-sector particles remain in thermal equilibrium with each other and share a common temperature $\Tp$. 

To determine the relic abundance of dark matter, we must solve the Boltzmann equations governing the evolution of the number densities of both the dark matter and the mediator particles. The hidden-sector temperature $\Tp$ enters these equations through the equilibrium distributions and the thermally averaged cross sections, for which the relevant energy scale is 
$\Tp$. Therefore, in order to solve the evolution equations consistently, we must first determine $\Tp$. A convenient way to do so is by extracting it from the hidden-sector energy density $\rho^\prime$. Since the population inside the hidden sector is much smaller than inside the visible sector, $\rho^\prime$ can be obtained by calculating the amount of energy that is transferred by scattering of SM particles into dark sector particles. The backreaction, i.e.~scattering of hidden sector particles into SM particles, can be neglected, as $\Tp \ll T$ and hence $n_{\mathrm{HS}}^{\mathrm{eq}}(\Tp) \ll n_{\mathrm{SM}}^{\mathrm{eq}}(T)$. By integrating the Boltzmann equations for the phase space density $f(p,T)$ convoluted with the particles' energy and assuming isotropy and homogeneity, we obtain a Boltzmann equation for the evolution of the hidden sector energy density, $\rho^\prime$,
\begin{align}
\label{eq:generalETBE}
    \frac{\partial\rho'}{\partial t}+3H\,\left(\rho'+P'\right)=&\int\frac{d^3 p}{(2\pi)^3}C[f(p,t)]
\end{align}
where $C[f]$ is the collision operator which encodes the information on the interactions of the hidden sector particles with the SM particles. By performing a change of variables, Eq.~\eqref{eq:generalETBE} can be rewritten as
\begin{equation}\label{eq:Tprime}
\frac{d\rho^\prime}{d T^\prime} \frac{d T^\prime}{d z}\frac{d z}{d t}=-3H(\rho^\prime+P^\prime)+\int\frac{d^3 p}{(2\pi)^3}C[f(p,t)] \;,
\end{equation}
with $z=m_\chi/T$. 
In our case, energy can be transferred via $2$ to $2$ scattering processes of SM fermions into DM, $f\bar{f} \rightarrow \chi \bar{\chi}$, mediated by the scalar or the pseudoscalar; by 2 to 2 scattering processes of SM particles into mediator particles of the form $i\; j \rightarrow a \; k$ and $i\; j \rightarrow \phi \; k$, where $i,j,k$ denote SM particles; and by inverse decays of SM fermions into mediator particles $f\bar{f} \rightarrow a$ and $f\bar{f} \rightarrow \phi$. 
The integrated collision term takes different expressions and can considerably be simplified, depending on the specific process that transfers energy to the hidden sector. We list below the expressions for the three cases that are important for our analysis \cite{Chu:2011be,Hambye:2019dwd,Bharucha:2022lty}:
\begin{itemize}
\item For equal masses of the initial state particles, $m_1=m_2=m$ we have for inverse decays $1\; 2 \rightarrow 3$
\begin{align}
\nonumber \int\frac{d^3 p_3}{(2\pi)^3}C[f_3]=&T\,\frac{g_1 g_2}{2 \pi^2}\, \Gamma_3\,m_3^3\,K_2\left(\frac{m_3}{T}\right)\Theta(m_3^2-4m^2) \; .
\end{align}
where $\Gamma_3$ denotes the decay width of particle 3. In our model, these include $f \bar{f} \rightarrow a$ and $f \bar{f} \rightarrow \phi$.
\item Apart from inverse decays, we only deal with $1\;2\rightarrow 3\;4$ processes, for which the integrated collision term can be expressed as
\begin{equation}
\int\frac{d^3 p}{(2\pi)^3}C[f]=g_1 g_2 \int \frac{d^3p_1}{(2\pi)^3}\frac{d^3p_2}{(2\pi)^3}f_1(p_1)f_2(p_2)v_\text{M\o l}\,\mathcal{E}(\vec{p}_1,\vec{p}_2) \;.
\label{eq:coll1}
\end{equation}
with  $v_{\text{M\o l}}$ the M\o ller velocity and $\cal E$  the energy transfer rate 
\begin{align}\hspace{-.4cm}
\label{eq:Entransfer}\mathcal{E}(\vec{p}_1,\vec{p}_2)=&\frac{1}{2 E_1 2E_2 v_\text{M\o l}}\int\prod_{i=3,4} \frac{d^3 p_i}{(2\pi)^3}\frac{1}{2 E_i} |\mathcal{M}|^2(2\pi)^4\delta^{(4)}(p_1+p_2-p_3-p_4)\Delta E_{tr}\,.
\end{align}
$\mathcal{M}$ is the matrix element for the $1\;2\rightarrow 3\;4$ process and $\Delta E_{tr}$ denotes the transferred energy. For final states containing two hidden sector particles with identical masses, the expression of this integrated collision term can be considerably simplified and requires only a single integral in the CM energy squared \cite{Chu:2011be, Hambye:2019dwd},
\begin{align}\label{eq:ETBE-chichi}
\int\frac{d^3 p_3}{(2\pi)^3}C[f_3]=&\frac{g_1 g_2}{32 \pi^4} \int ds\, \sigma(s)\,(s-4 m^2) \,s\, T K_2\left(\frac{\sqrt{s}}{T}\right)
\end{align}
with the limits of integration given in Refs.~\cite{Gondolo:1990dk,Chu:2011be}. 
This expression can be applied to the energy transfer via the process $f\bar{f}\rightarrow \chi \bar{\chi}$.
\item For energy transfer from the SM to the \emph{mediator particles} via $1\;2\rightarrow 3\;4$ processes, we are however dealing with cases $i \;j \rightarrow a \; k$ and $i\; j \rightarrow \phi \; k$ where $i$, $j$, $k$ denote SM particles (see Sec.\ref{sec:model}). Therefore, only a certain fraction of the final state energy is transferred to the hidden sector. In this case, the collision term can be expressed as
\begin{align}\label{eq:Etransfer}
\mathcal{E}(\vec{p}_1,\vec{p}_2)=\frac{1}{8\pi F}\int p_3^2d p_3dc_{13}\frac{1}{4 E_4} |i\mathcal{M}|^2\frac{\delta(p_3-p_3^0)}{|g'(p_3)|_{p_3\to p_3^0}}\;,
\end{align}
with ``$3$''  denoting again the particle in the hidden sector (for a detailed derivation see App.~B of~\cite{Bharucha:2022lty}).
Here, $g(p_3)=E_1+E_2-E_3-E_4$ is the argument of the delta-function in Eq.~\eqref{eq:Entransfer}, $c_{13}\equiv \cos (\theta_{13})$ and $\theta_{13}$ is the angle between the momenta of particles 1 and 3. This expression is then inserted into Eq.~\eqref{eq:coll1} and the integration is performed numerically, where we make the approximation that $E_3\approx p_3$. 

\end{itemize}
Solving Eq.~\eqref{eq:generalETBE} with the corresponding expressions for the right-hand side enables us to obtain $\rho'$ as a function of the photon temperature $T$.
The energy density of the hidden sector is related to its temperature $\Tp$ via the equation of state. When $\Tp \gg m_a, m_\phi, m_\chi$ the hidden sector particles are ultra relativistic. As long as $\Tp > m_a, m_\phi, m_\chi$, they remain in thermal equilibrium. Finally, when $\Tp\lesssim m_i$ particle $i$ begins to decouple. Consequently, depending on the value of $\Tp$, we adopt different forms for the equation of state, which split into three regimes: 

\begin{itemize}
\item \textbf{Initial condition ($\Tp \gg m_a, m_\phi, m_\chi$):} At initial times, the energy density of the universe will be dominated by that of the SM particles, $\rho\,\propto\, T^4$. When sufficient energy has been transferred to the hidden sector, all hidden sector particles are ultra-relativistic, and the equation of state is given by:
\begin{equation}
P^\prime=\rho^\prime/3 \qquad \text{for } \Tp \gg m_i, \quad i=a,\phi,\chi \;,
\end{equation}
allowing us to rewrite Eq.~\eqref{eq:generalETBE} as 
\begin{equation}\label{eq:EBEinT}
 \frac{\partial}{\partial T}\left(\frac{\rho'}{\rho}\right)=-\frac{1}{HT\rho}\int\frac{d^3 p}{(2\pi)^3}C[f(p,t)]\,.
\end{equation}
Integrating this equation serves as an initial condition for 
$\rho'/\rho$. Setting
\begin{equation}\label{eq:rhoprimeoverrho}
\left(\frac{\rho^\prime}{\rho}\right)_0=\frac{\sum_{i=\chi,\bar{\chi},a,\phi}\rho_i^{\rm eq}(\Tp_0)}{\frac{\pi^2}{30}g_{\rm eff,SM}(T_0)T_0^4} \; .
\end{equation} 
with $\rho_i^{\rm eq}$ the Maxwell-Boltzmann equilibrium distribution, we obtain the initial $\Tp_{0}$.

\item \textbf{Equilibrium ($\Tp > m_a, m_\phi, m_\chi$):} Once thermal equilibrium is achieved among the hidden sector particles, this state is maintained at a common temperature $\Tp$ for $\Tp > m_\chi,m_a,m_\phi$. 
All hidden sector particles follow their Maxwell-Boltzmann equilibrium distributions at temperature $\Tp$ and the hidden sector equation of state is simply given by the sum of these equilibrium values 
\begin{equation}\label{eq:eqStateEquilibrium}
\rho^\prime+P^\prime=\sum_{i=a,\phi,\chi,\bar{\chi}}\rho^{\rm{eq}}_i(\Tp)+P^{\rm{eq}}_{i}(\Tp) \; ,
\end{equation}
which can then be inserted in Eq.~\eqref{eq:Tprime}.

\item \textbf{Decoupling ($\Tp\lesssim m_i$):} Once $\Tp$ drops below the mass of a hidden sector particle, $\Tp\lesssim m_i$, its equilibrium distribution becomes Boltzmann suppressed. Therefore we allow for the possibility that the particles' phase space distribution will start to diverge from the equilibrium distribution, as will consequently the energy density and the pressure:
\begin{equation}\label{eq:eqStateDecoupling}
\rho_i = \frac{\rho_{i}^{\rm{eq}}(\Tp)}{n_{i}^{\rm{eq}}(\Tp)} \ n_i\,, \qquad
P_i = \frac{P_{i}^{\rm{eq}}(\Tp) }{n_{i}^{\rm{eq}}(\Tp) }\ n_i=T^\prime n_i \; .
\end{equation}
 Here we made the assumption that the DM phase space distribution is proportional to the Maxwell-Boltzmann distribution, and that the proportionality factor is a function of temperature only.
\end{itemize}

Having determined the hidden-sector temperature $\Tp$ and specified the corresponding equation of state, we can now turn to the dynamical evolution of the full coupled system, in which $\Tp$ evolves alongside the number densities of the hidden-sector particles. 
In analogy to the hidden sector, the SM particles are assumed to efficiently thermalize among themselves and remain in kinetic equilibrium with the photon bath at temperature  $T$. 
 For particles in equilibrium at a given temperature, whether $T$ or $\Tp$, their distributions can therefore be replaced by the corresponding equilibrium distributions. 
Here we neglect quantum statistical factors and assume Maxwell-Boltzmann statistics, $f_{\rm eq}(p,T)=\exp(-E/T)$.
For particles that are also in chemical equilibrium, we can further apply the principle of detailed balance, $\braket{\sigma_{i j \rightarrow k l} v}n_{i}^{\rm{eq}} n_{j}^{\rm{eq}}=\braket{\sigma_{k l \rightarrow i j} v} n_{k}^{\rm{eq}} n_{l}^{\rm{eq}}$, with $\braket{\sigma_{i j \rightarrow k l} v}$ the thermally averaged cross section for the $i\; j\rightarrow k\; l$ process and $n_i^{\rm{eq}}$ is the Maxwell-Boltzmann equilibrium number density of particle species $i$ (below we define $n_i^{\rm{eq}}\equiv n_i^{\rm{eq}}(T)$).

Given these simplifications, the general set of Boltzmann equations governing the evolution of the $a$, $\phi$ and $\chi$ number densities is then given by
\begin{align}
    \frac{\mathrm{d} n_\chi}{\mathrm{d}t} + 3 H n_\chi
    &=
    \sum_f \ev*{\sigma_{\chi\bar \chi\to f \bar f}v}(T) \big[ (n_\chi^{\text{eq}}) ^2 - n_\chi^{2} \big]
        + \ev*{\sigma_{a\phi \to \chi\bar\chi} v}(\Tp) n_a n_\phi
    \nonumber\\
    &\quad
    - \ev*{\sigma_{\chi\bar\chi \to a\phi} v}(\Tp) n_\chi^2
    + \ev*{\sigma_{aa \to \chi\bar\chi} v}(\Tp) n_a^2
    - \ev*{\sigma_{\chi\bar\chi \to aa} v}(\Tp) n_\chi^2
    \nonumber\\
    &\quad
    + \ev*{\sigma_{\phi\phi \to \chi\bar\chi} v}(\Tp) n_\phi^2
    - \ev*{\sigma_{\chi\bar\chi \to \phi\phi} v}(\Tp) n_\chi^2\,,
    \nonumber\\
    \frac{\mathrm{d} n_a}{\mathrm{d} t} + 3 H n_a
    &=
    \sum_{i, j, k} \ev*{\sigma_{i a \to j k} v}(T) \big( n_a^{\text{eq}} n_i^{\text{eq}} - n_a n_i^{\text{eq}} \big)
    + \ev*{\Gamma_{a \to f \bar f}}(T) \big( n_a^{\text{eq}}   - n_a \big)
    \nonumber\\
    &\quad
    - \ev*{\sigma_{aa \to \chi\bar\chi} v}(\Tp) n_a^2
    + \ev*{\sigma_{\chi\bar\chi \to aa} v}(\Tp) n_\chi^2 
    \nonumber\\
     &\quad  
     - \ev*{\sigma_{a\phi \to \chi\bar\chi} v}(\Tp) n_a n_\phi
    + \ev*{\sigma_{\chi\bar\chi \to a\phi} v}(\Tp) n_\chi^2 \,,
    \nonumber\\
    \frac{\mathrm{d} n_\phi}{\mathrm{d} t} + 3 H n_\phi
    &=
    \sum_{i, j, k} \ev*{\sigma_{i \phi \to j k} v}(T) \big( n_\phi^{\text{eq}} n_i^{\text{eq}} - n_\phi n_i^{\text{eq}} \big)
    + \ev*{\Gamma_{\phi\to f\bar f}} \big( n_\phi^{\text{eq}} - n_\phi \big)
    \nonumber\\
    &\quad
    - \ev*{\sigma_{\phi\phi \to \chi\bar\chi} v}(\Tp) n_\phi^2
    + \ev*{\sigma_{\chi\bar\chi \to \phi\phi} v}(\Tp) n_\chi^2
    \nonumber\\
    &\quad   - \ev*{\sigma_{a\phi \to \chi\bar\chi} v}(\Tp) n_a n_\phi
    + \ev*{\sigma_{\chi\bar\chi \to a\phi} v}(\Tp) n_\chi^2 \, ,
    \label{eq:generalBoltzmann}
\end{align}
with the thermally averaged decay rate of particle $i$ 
\begin{equation}\label{eq:Gammav}
\braket{\Gamma_i}=\Gamma_i \frac{K_1(m_i/T)}{K_2(m_i/T)} \; ,
\end{equation}
and the thermally averaged cross section for the $1\; 2 \to 3\; 4$ process \cite{Gondolo:1990dk}
\begin{equation}\label{eq:sigmav}
\braket{\sigma_{12\to 34}v}=\frac{C}{2\,T K_2(m_1/T)\,K_2(m_2/T)}\int_{s_{\rm{min}}}^\infty \sigma(s) \frac{F(m_1,m_2,s)^2}{m_1^2 m_2^2\sqrt{s}}\,K_1(\sqrt{s}/T)\, d s\,\,.
\end{equation}
Here, $\Gamma_i$ is the decay width of particle $i$ and $\sigma(s)$ is the 2 to 2 cross section as a function of the centre-of-mass energy squared and 
\begin{align}
F(m_1,m_2,s)=\frac{\sqrt{(s-(m_1+m_2)^2)(s-(m_1-m_2)^2)}}{2} \;.
\end{align} 
The lower limit of the integral in \eqref{eq:sigmav} is $s_{\rm{min}}= \text{max}\left((m_1+m_2)^2,(m_3+m_4)^2\right)\,$. Furthermore, for processes with two identical particles in the initial state, we need to include an additional factor of $1/2$, since the phase-space integral over-counts the configurations of identical particles. This is taken care of by the constant $C$ which is $1/2$ $(1)$ for identical (non-identical) initial state particles, respectively. 
In Eqs.~\eqref{eq:generalBoltzmann} we have made explicit the dependence of the thermally averaged cross sections and decay rates on the temperature at which the processes happen, since the hidden sector temperature $\Tp \neq T$. As explained above, this hence couples Eqs.~\eqref{eq:generalBoltzmann} with Eq.~\eqref{eq:Tprime}. The rate of expansion of the universe is given by the Hubble function $H = \left( \frac{8}{3} \pi G \rho \right)^{1/2}$
with $G$ the gravitational constant and $\rho=\rho_{\rm{SM}}+\rho_{\rm{HS}}$ the total energy density of the universe. 
The energy density of the SM particles is $\rho_{\rm{SM}} = g_{*\rho,\rm{SM}}(T)T^4\pi^2/30$, 
where $g_{*\rho,\rm{SM}}(T)$ are the SM effective degrees of freedom in energy, for which we use the fit functions provided in~\cite{Saikawa:2018rcs}. 
The explicit form for the hidden sector energy density is given by Eqs.~\eqref{eq:eqStateEquilibrium} or \eqref{eq:eqStateDecoupling}, depending on the value of $\Tp$. 
The relevant DM, scalar and peudoscalar number density changing processes 
in the set of coupled Boltzmann equations \eqref{eq:generalBoltzmann} include (pseudo-) scalar-mediated interactions between SM fermions and DM ($f\bar f \leftrightarrow \chi \bar \chi$), the HS interactions ($a a \leftrightarrow \chi \bar \chi $, $\phi \phi \leftrightarrow \chi \bar \chi $, $\chi \bar \chi \leftrightarrow a\phi$, $aa \leftrightarrow \phi$) and finally the SM-mediator interaction processes ($i a \leftrightarrow j k$, $i \phi \leftrightarrow j k$, with $i$, $j$, $k$ SM particles). The relevant scale for the HS interaction processes is the hidden sector temperature $\Tp$.  
We have not included any higher-dimensional operators in the effective Lagrangian in Eq.~\eqref{eq:Lag}. 
Although these operators are UV-dominated and generally introduce a dependence on the reheating temperature for freeze-in processes, this dependence is very mild for the DFO mechanism \cite{Bharucha:2022lty}. 
We hence neglect these operators here, and the dark matter relic density is IR dominated \footnote{Likewise, we neglect finite temperature effects as is usually done for freeze-out scenarios.  Such effects can be important for freeze-in which occur at higher temperatures, see for example~\cite{Belanger:2020npe}.}. 
We choose $T_{RH}=10^4$ GeV as the value of the reheating temperature in our calculations. 

\section{Constraints}\label{sec:constraints}
The unique structure of DFO leads to an equally distinctive landscape of experimental constraints. While the feeble portal couplings to the SM place DFO within the broad class of non-thermal production mechanisms, its phenomenology offers a more nuanced picture with an underappreciated hybrid featuring feeble coupling to the visible sector alongside strong interactions within the dark sector that drive internal thermalisation. This combination gives rise to a distinctive phenomenology. On the one hand, non-thermal features, such as late decays of long-lived mediators that can impact cosmological observations (e.g.~the synthesis of light elements during BBN). On the other hand, effective dark sector annihilations, particularly when $s$-wave, can produce signals in the early universe or astrophysical settings like galaxies, akin to thermal freeze-out. While such annihilation-induced signatures do require $s$-wave processes to remain visible today,  the key distinguishing feature of DFO lies in the coexistence of these non-thermal and thermal-like effects. As a result, DFO opens up rich multimessenger detection prospects, while simultaneously evading conventional CMB and direct detection constraints in the sub–10 GeV range, and potentially imprinting testable signatures on BBN. We enumerate the constraints in the following subsections.

\subsection{CMB anisotropies}
\label{subsec:cmb}
Annihilation of DM particles during the recombination epoch can inject energy into the primordial plasma in the form of electromagnetically interacting particles such as electrons, positrons, and photons. This injection increases the residual ionisation fraction and alters the thermal and ionisation history of the universe, thereby modifying the temperature and polarisation anisotropies of the cosmic microwave background (CMB). Since CMB anisotropies have been precisely measured, particularly by the \texttt{Planck} satellite, such energy injection is tightly constrained. In particular, for annihilation channels into $e^+\,e^-$ pairs, \texttt{Planck} excludes the canonical thermal relic cross section $\langle\sigma v\rangle_{\rm th}\approx 3\times 10^{-26}\,{\rm cm^3/s}$ for DM masses below roughly 10 GeV. These bounds severely limit $s$-wave annihilation scenarios during recombination. 

However, decoupled freeze-out scenarios can naturally evade CMB constraints, even for $s$-wave annihilation. This is primarily because thermal equilibrium is established within the dark sector at a lower temperature, $T' \lesssim T$, compared to the visible sector. As a result, a smaller dark-sector coupling is sufficient to achieve the required $\langle \sigma v \rangle_{\rm th}$ for freeze-out. The weak portal coupling to SM, e.g.~$g_{a(\phi)f}$, controls the efficiency of energy transfer between sectors and effectively sets the value of $T'$. This opens up a broader region of relic density-compatible parameter space, typically forming a band in the $g_{a(\phi)\chi}$ vs.~$m_\chi$ plane, as discussed in Sec.~\ref{sec:results}. We highlight this feature for the first time in the context of DFO as an effective workaround for reviving light $s$-wave annihilation scenarios; on par with alternative mechanisms such as Breit-Wigner resonance~\cite{Belanger:2025kce,Binder:2022pmf}.

CMB measurements set an upper limit on the thermal-averaged annihilation cross section given by: 

\begin{align}
\langle \sigma v \rangle \lesssim \frac{p_{\rm ann} \, m_\chi}{\sum_i f^i_{\rm eff} \, {\rm Br}_i}
\label{eq:cmb}
\end{align}
where $p_{\rm ann} = 3.2 \times 10^{-28}\ {\rm cm^3\,s^{-1}\,GeV^{-1}}$~\cite{Planck:2018vyg}, $f^i_{\rm eff}$~\cite{Slatyer:2015jla} denotes the energy injection efficiency for the $i$-th annihilation channel, and ${\rm Br}_i$ is the corresponding branching ratio. We use \texttt{micrOMEGAs} to obtain the CMB constraint~\cite{Alguero:2023zol}. Since the mediators are extremely weakly coupled to the SM, the direct $2 \to 2$ annihilation channels are highly suppressed. Instead, the dominant contribution in $\langle \sigma v\rangle$ comes from $2 \to 4$ processes, where DM annihilates into a pair of on-shell mediators, each subsequently decaying into a pair of SM particles. Among the dark sector annihilation channels, $\chi \bar{\chi} \to s\, a$ is the only $s$-wave process and therefore dominates the annihilation branching ratios at the time of recombination. As a result, the CMB constraint translates into an upper bound on the dark sector coupling, which scales with the cross section as $g_{a\chi} = g_{\phi\chi} \propto \left( \langle \sigma v \rangle \right)^{1/4}$.

\subsection{Indirect detection}
\label{subsec:id}
Dark sector annihilation not only constrains the parameter space during recombination but also offers strong potential for generating electromagnetic excesses in galaxies, clusters, and dwarf galaxies at the present epoch. Several studies in the literature have highlighted the potential of $2 \to 4$ annihilation channels for $\gamma$-ray searches in galaxies and dwarf spheroidals~\cite{Profumo:2017obk,Pospelov:2007mp,Su:2025mxv,Datta:2023ncp}, as well as for neutrino observations~\cite{Du:2020avz, ANTARES:2022aoa}. However, significant room remains to explore uncharted territory, including possible signatures in X-ray and radio observations. These alternative channels are particularly relevant in scenarios involving light mediators and cascade decays, where low-energy electrons and positrons can give rise to detectable synchrotron or inverse Compton signals. 

\subsubsection{Radio constraints}
\label{subsec:radio}

DM annihilation into electrons and muons can produce synchrotron radiation as the resulting relativistic $e^\pm$ interact with magnetic fields in galaxy clusters and dwarf galaxies. Current radio telescopes such as the Green Bank Telescope~\cite{Natarajan:2015hma}, MeerKAT~\cite{Booth:2009ex}, and others are sensitive to this emission, offering a complementary probe of DM annihilation. In this work, we incorporate recent radio constraints from \texttt{MeerKAT} telescope, a precursor to the upcoming SKA collaboration~\cite{Cembranos:2019noa}, and focus on radio emissions from galaxy clusters, which offer more robust limits compared to dwarf spheroidal galaxies, due to reduced astrophysical uncertainties~\cite{Lavis:2023jju}. Clusters benefit from better-constrained modeling parameters, particularly due to their $\mu$G-scale magnetic fields and reduced uncertainties from diffusion effects. As a result, the upper limits on the DM annihilation cross section are significantly less sensitive to astrophysical systematics and variations in halo or diffusion parameters.

We use \texttt{DarkMatters}~\cite{Sarkis:2024zjg} to compute the synchrotron flux from DM annihilation, where the evolution of $e^\pm$ is driven by interactions with the ambient magnetic field and thermal gas, both of which are abundant in large-scale structures such as galaxy clusters. These interactions are modeled using a cosmic-ray transport equation that captures the effects of spatial diffusion and energy losses, this equation is solved through the full diffusion-loss equation. The latter includes a {\it source function} $Q_{\rm e}$, which encodes the model-specific injection of $e^\pm$ from DM annihilation~\cite{Lavis:2023jju,Beck:2023oza}.
\begin{align}
\frac{\partial}{\partial t} \left( \frac{dn_{\rm e}}{dE} \right) = \vec{\nabla} \cdot \left[ D(E,\vec{r})\, \vec{\nabla} \left( \frac{dn_{\rm e}}{dE} \right) \right] + \frac{\partial}{\partial E} \left[ b(E,\vec{r})\, \frac{dn_{\rm e}}{dE} \right] + Q_{\rm e}(E,\vec{r}) \; , \label{eq:diff-loss}
\end{align}
where \( D(E,\vec{r}) \) denotes the diffusion coefficient, \( b(E,\vec{r}) \) the energy-loss rate, and \( dn_{\rm e}/dE \) the electron distribution. The source term contains the model dependence for $e^\pm$ injection, and the contribution from the \( i \)-th annihilation channel, \( Q_{{\rm e},i}(r,E) \), takes the form:
\begin{align}
Q_{{\rm e},i}(r,E) = \frac{1}{2} \langle \sigma v \rangle \left( \frac{dN^ i_{\rm e}}{dE} \right) \left( \frac{\rho_\chi(r)}{m_\chi} \right)^2 \; ,
\end{align}
where \( \rho_\chi(r) \) is the DM density profile, and \( i \) labels a specific annihilation channel.

The above equation can be solved using the Green's function method to determine the synchrotron emissivity, which in turn allows the computation of observable quantities such as the radio surface brightness and flux density spectrum. These are then compared against \texttt{MeerKAT}'s L-band sensitivity~\cite{Lavis:2023jju}, based on the \texttt{MeerKAT} Galaxy Cluster Legacy Survey (MGCLS)~\cite{Knowles:2021nvz}, which includes approximately 1000 hours of L-band (900-1670 MHz) observations of 115 galaxy clusters.

In this work, we focus on the cluster {\it Abell 133} for deriving the relevant radio constraints, owing to its cuspy NFW-like density profile and favourable magnetic field configuration. These features result in limits that surpass existing \texttt{Fermi-LAT} $\gamma$-ray constraints over a wide region of parameter space, as can be seen in Fig.~\ref{fig:constraints}.

\subsubsection{\texorpdfstring{$\gamma$}{gamma}-ray searches from dwarf galaxies}
\label{subsec:fermilat}

DM annihilation into SM final states can also produce photons, either directly or via decay and radiation. In regions of high DM density, such as dwarf spheroidal galaxies(dSphs), this photon flux can be detected by instruments like the Fermi Large Area Telescope (\texttt{Fermi-LAT}), which observes $\gamma$-ray emission from the Milky Way's dSPhs. These galaxies are ideal targets for such searches due to their proximity, high DM content, and minimal astrophysical backgrounds. 

\texttt{Fermi-LAT}'s non-observation of any significant excess consistent with DM annihilation has led to stringent constraints on $\langle \sigma v \rangle$~\cite{Fermi-LAT:2015att}. The differential $\gamma$-ray flux from DM annihilation in a solid angle $\Delta\Omega$ is given by~\cite{Bergstrom:1997fj}:

\begin{align}
\frac{d\Phi_\gamma}{dE}(\Delta\Omega) = \frac{1}{4\pi \eta} \frac{\langle \sigma v \rangle}{m_\chi^2} \left( \frac{dN_\gamma}{dE} \right) \cdot J,
\end{align}
where $\eta = 2$ and $\eta = 4$ for self-conjugate and non self-conjugate DM, respectively; in our analysis, we consider the latter. $dN_\gamma/dE$ denotes the average photon spectrum per annihilation, and $J$ is the astrophysical $J$-factor, defined as:

\begin{align}
J = \int_{\Delta\Omega} d\Omega \int_{\text{l.o.s}} ds \, \rho_{\chi}^2(s),
\end{align}
where $\rho_{\chi}(s)$ is the DM density along the line of sight (l.o.s), typically modeled using an NFW profile~\cite{Navarro:1996gj}, as adopted in standard likelihood-based analyses.

In this work, we use \texttt{micrOMEGAs}~\cite{Alguero:2023zol}, which includes  the code~\cite{calore_2021_5592836} using a data-driven approach to estimate background and derive limits from \texttt{Fermi-LAT} based on  observations of 25 Dwarf galaxies~\cite{Alvarez:2020cmw,Calore:2018sdx} .

\subsection{Big Bang Nucleosynthesis}
After the freeze-out of DM its comoving number density remains effectively fixed. However, the $a$ and $\phi$ will eventually decay into SM states. Due to the fact that they couple feebly to SM particles, these decays occur relatively late and can be constrained by Big Bang Nucleosynthesis (BBN).

BBN refers to the early formation of light elements, such as deuterium, helium-3 and helium-4, when the CMB temperature cools to around 0.1 MeV, a point where photon scatterings no longer cause dissociation.   
Precise predictions of these abundances align well with astrophysical observations, imposing strict limits on beyond-the-Standard-Model particles, the presence of which would alter the abundances. 
BBN hence plays an important role in constraining the existence of new long-lived particles in the early universe. 
Depending on their mass, lifetime, and abundance, these new particles can have various effects on BBN predictions.  
For the mediator masses we consider here, constraints from photo- and hadrodissociation are relevant: When the mass of the new particle exceeds twice the energy needed to ionize helium-4 and deuterium, energetic photons from electromagnetic cascades can efficiently dissociate these nuclei. For slightly heavier particles, decay processes involving pions can also come into play. Pion-induced scattering can drive proton-neutron conversions, increasing the neutron-to-proton ratio and altering the synthesis of deuterium and heavier nuclei. 
At even higher masses, above a few GeV, decays into hadrons produce hadronic cascades of pions and nucleons that dissociate nuclei.
Importantly, such effects are mitigated if the particle has a very short lifetime ($\tau<10^{-2}$ s), as neutron-to-proton equilibrium can then be restored through weak interactions.

The scenarios from dissociation processes due to hadronic and electromagnetic interaction processes provide us with a bound on the lifetime of long-lived (sub)-GeV mediator particles, determined by the energy injected during their decay. Since the decay width and consequently the lifetime depend quadratically on the coupling to SM fermions, for sufficiently abundant mediator particles, BBN provides bounds on $g_{af}=g_{\phi f}$. The amount of injected energy during the decay depends in return on the mediators' masses and abundances. We apply the exclusion limits from~\cite{Kawasaki:2017bqm} for the case where the mediators decay into hadrons and from~\cite{Kawasaki:2020qxm} for electromagnetic decays. In these references, the photo- and hadrodissociation of light elements were calculated and compared to the latest observational data, providing upper bounds on the relic abundance of the decaying particle as a function of its lifetime.

%
\section{Results and observations}
\label{sec:results}

To study qualitatively different scenarios, we choose three masses for the mediator particles and assume the masses of the scalar and pseudoscalar mediators to be identical. We will hence consider the cases (1) $m_a=m_\phi=250$ MeV, (2) $m_a=m_\phi=3$ GeV and (3) $m_a=m_\phi=30$ GeV. The first case is chosen such that hadronic decays are not kinematically allowed and the mediator particles decay almost exclusively into muons. In the second case the mediators are heavy enough such that they decay hadronically into quarks and gluons rather than into mesons. The third case illustrates the case of a weak scale mediator.  We explore the parameter space for $m_\chi=5\ {\rm GeV}$ to $m_\chi=M_W$ ($W$ boson mass) and impose the additional condition that $m_\chi> m_a / m_\phi$ to guarantee that the DM mainly annihilates into mediators via $\chi\chi\to a\phi$.

%
%
%

As explained in Sec.~\ref{sec:DFO}, in order to obtain the DM relic density for a given set of couplings and masses, a set of four coupled differential equations, namely Eqs.~\eqref{eq:generalBoltzmann} together with Eq.~\eqref{eq:generalETBE}, has to be solved. We solve these equations numerically,  until the comoving DM number density $Y_{\text{DM}}=(n_\chi+n_{\bar{\chi}})/s$ stays constant. To do so, we adapted the code that was developed for~\cite{Bharucha:2022lty}. Given the complexity of scanning the entire parameter space of the model defined in Eq.~\eqref{eq:LagDS-ferm}, we restrict our analysis to the subspace defined by $g_{a\chi} \approx g_{\phi\chi}$ and $g_{af} \approx g_{\phi f}$. Relaxations of these assumptions are discussed below. 

 The parameter space in the $g_{af}=g_{\phi f}$, $m_\chi$ plane is bounded by two conditions: (1) the very definition of a ``decoupled freeze-out'', i.e.~maintaining a \emph{decoupled} hidden sector, and (2) ensuring sufficient energy transfer to the hidden sector to produce the observed DM relic density. These conditions are independent of the hidden sector interactions and therefore independent of $g_{a\chi}=g_{\phi\chi}$. Based on these conditions, a three-dimensional grid of values $m_\chi$, $g_{a(\phi)f}$ is constructed to map plausible mediator-DM couplings $g_{a(\phi)\chi}$.For each of these parameters, we then solve the coupled Boltzmann equations \eqref{eq:generalBoltzmann} and \eqref{eq:generalETBE}. The correct coupling $g_{a(\phi)\chi}$ is determined using a bisection method to find the value that yields the DM relic density observed today, namely \cite{Planck:2018vyg}
 $$
\Omega_{\mathrm{DM}} h^2=\frac{\rho_{\mathrm{DM}}}{\rho_{\text {crit }} / h^2}=\frac{m_\chi (2 \cdot Y_{\chi}) s_0}{\rho_{\text {crit }} / h^2}=0.120(1) 
 $$
 where $\rho_{\text {crit }} / h^2=1.053672(24) \times 10^{-5} \mathrm{GeV} \mathrm{~cm}$ and $s_0=2891.2 \mathrm{~cm}^{-3}$. 

\begin{figure}
    \centering
    \includegraphics[width=\linewidth]{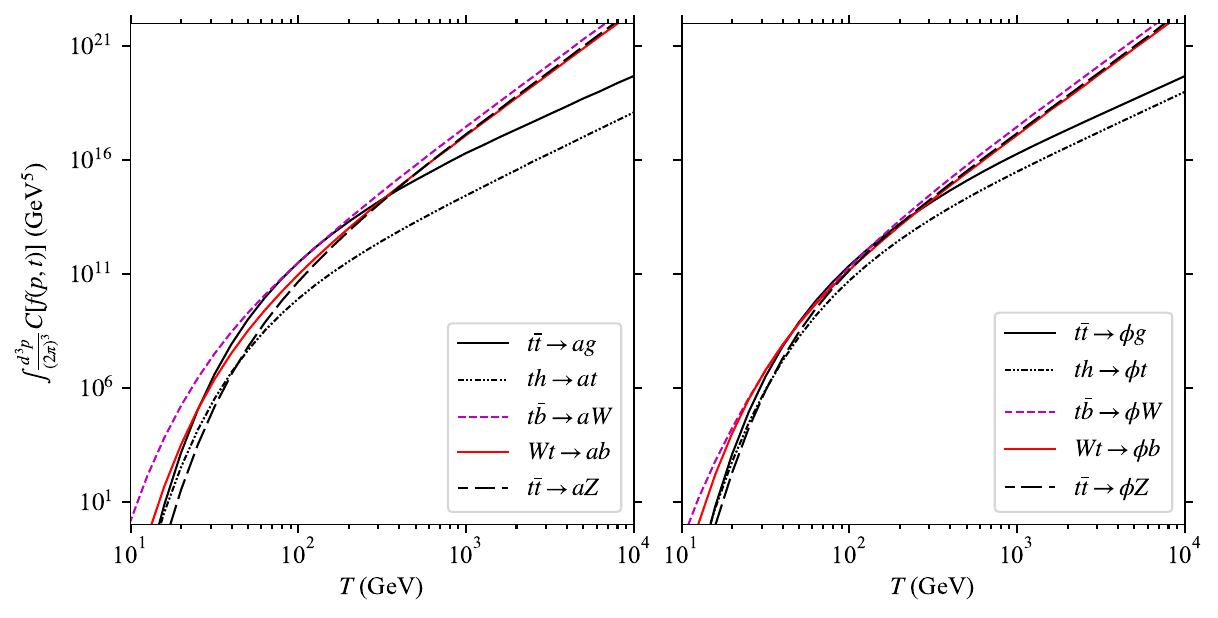}
    \caption{Integrated collision term for a selection of the most dominant processes $i\; j \rightarrow a \; k$ (left) and $i\; j \rightarrow \phi \; k$ (right) for the energy transfer Boltzmann equation \eqref{eq:Entransfer}, where $i,j,k$ denote SM particles, as a function of the photon temperature $T$. We set $g_{af}=g_{\phi f}=1$ GeV$^{-1}$. }
    \label{fig:e-transfer}
\end{figure}

The solution of the collision term for the energy transfer in Eq.~\eqref{eq:generalETBE} for the cases $i\; j \rightarrow a\; k$ and $i\; j \rightarrow \phi\; k$, with $i$, $j$, $k$ SM particles involves evaluating a high-dimensional integral, see Eq.~\eqref{eq:Etransfer}. 
Since the SM-mediator coupling is proportional to the mass of the fermionic SM particle, processes involving the top or the bottom quark will be dominant. 
We include the dominant processes $Vf \rightarrow af^\prime$, $f\bar f^{(\prime)}\rightarrow a V$, $fj\rightarrow a f$, $f \bar f\rightarrow a j$ where $V=\{W,Z\}$, $j=\{g,h\}$, $f=\{t,b\}$, and $f^\prime$ denotes the isospin partner of $f$, along with their scalar counterparts $(a\leftrightarrow \phi)$ and hermitian conjugates where applicable.
We include the same selection of processes in the Boltzmann Eqs.~\eqref{eq:generalBoltzmann}.  We plot the evolution of the right-hand side of the energy transfer Boltzmann equation \eqref{eq:Entransfer} for a selection of these processes as a function of the photon temperature in Fig.~\ref{fig:e-transfer} for illustrative purposes, where we have normalized the right-hand side by setting $g_{af}=g_{\phi f}=1$ GeV$^{-1}$.
We observe that processes involving electroweak bosons dominate the energy transfer, specifically $t\bar{t} \rightarrow aZ$, $Wt \rightarrow ab$, $t\bar{b} \rightarrow aW$, $tZ \rightarrow at$ and $Wb \rightarrow at$, along with the $(a\leftrightarrow \phi)$ counterparts and hermitian conjugates. At temperatures
below $\max (m_{i})$ the integrand is exponentially (Boltzmann) suppressed. Given that the processes are IR dominated, the dominant contribution arises from the saddle point of the integrand, which is located at energy scales set by the heaviest particle participating in the reaction, here the top quark. Therefore, $t\bar b\rightarrow a (\phi) W$ gives the dominant contribution to the energy transfer. 
 
\begin{figure}
    \centering
    \includegraphics[width=\linewidth]{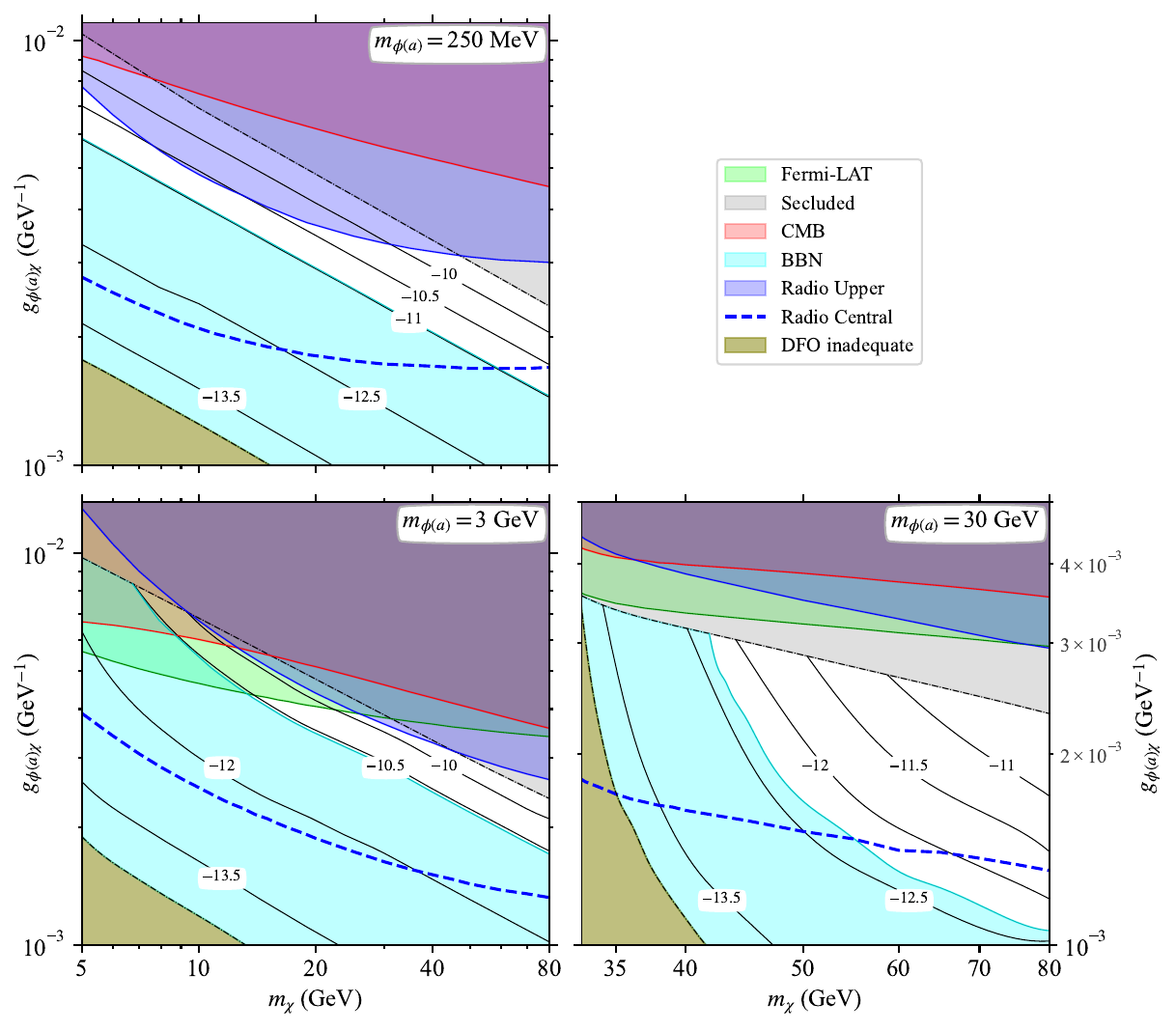}
     \caption{Decoupled freeze-out region in the  $(m_\chi,g_{a(\phi)\chi})$-plane for the three mediator masses $m_a=m_\phi= \{0.25, 3,30\}$ GeV, together with constraints from CMB, indirect detection and BBN. Black contour lines show the values for the SM-mediator couplings $g_{af}=g_{\phi f}$ that yield the observed DM relic density; the contour labels indicate $\log_{10}(g_{a(\phi)f})$. In the khaki region, there is an underproduction of DM while in the grey region the dark sector and the SM sector equilibrate.  The upper black dashed line corresponds to the combination of couplings in the secluded scenario. 
     Shaded regions show excluded parameter space, as detailed in the main text: CMB power spectrum (red), BBN (turquoise), and indirect detection constraints—including \texttt{Fermi-LAT} (green) and radio emission bounds (blue). For the radio constraints, the solid and dashed lines denote the upper error band and central value, respectively. The BBN exclusion contours shown correspond to a fixed value of $g_{\phi (a)f}$; they are largely insensitive to the value of $g_{\phi(a)\chi}$. Thus, each exclusion line represents the BBN bound associated with that specific $g_{\phi (a)f}$ contour. The white region represents allowed parameter space for decoupled freeze out.}
    \label{fig:constraints}
\end{figure}

After calculating the energy transfer and determining the hidden-sector temperature, we solve the Boltzmann equations to obtain the relic density in the DFO region. 
The results are illustrated in Fig.~\ref{fig:constraints}, for three fixed values of the mediator mass: $m_a=m_\phi=250$ MeV (left panel), $m_a=m_\phi=3$ GeV (middle panel) and $m_a=m_\phi=30$ GeV (right panel). Each plot shows the DM-mediator couplings, $g_{a\chi}=g_{\phi \chi}$, as a function of the DM mass, $m_\chi$. 
Black contours are shown for fixed values of $g_{af}=g_{\phi f}$ that reproduce the observed relic density for given $g_{a\chi}=g_{\phi \chi}$ and $m_\chi$, derived from the relic density analysis described earlier. 
For larger hidden sector couplings $g_{a\chi}=g_{\phi \chi}$, DM remains in equilibrium longer and decouples later, resulting in a smaller relic density (this is simply the fact that the relic density from freeze-out is inversely proportional to the coupling strength). To offset this effect, the SM-mediator couplings must increase to transfer more energy into the hidden sector initially, ensuring a larger DM equilibrium number density, $n_{\rm{eq}}(\Tp)$. Consequently, higher $g_{a\chi}=g_{\phi \chi}$ values correlate with larger $g_{af}=g_{\phi f}$. The dominant interaction inside the hidden sector for $g_{a\chi} = g_{\phi\chi} $ is the $s$-wave enhanced process $a \phi \leftrightarrow \chi \bar{\chi}$ and the relic density therefore approximately inversely proportional to $(g_{a\chi}g_{\phi\chi})^2$.
In the grey and khaki regions it is no longer possible to obtain the observed relic density from DFO: 
Above the upper boundary of the viable region, shown in grey, the couplings $g_{a(\phi)f}$ become large enough to bring the hidden sector into thermal equilibrium with the SM, thereby violating the decoupling condition. 
For instance, for $m_\chi=250$ MeV, once $g_{a(\phi)f}\gtrsim 2.5\times 10^{-10}$ GeV$^{-1}$ the hidden and visible sectors thermalize, marking the transition to the secluded regime. The upper boundary (dashed-dotted line) indicates the corresponding value of $g_{a(\phi)\chi}$ required to obtain the correct relic abundance in this \emph{secluded regime}. This relation is valid for all $g_{a(\phi)f}\gtrsim 2.5\times 10^{-10}$ GeV$^{-1}$, where the relic density becomes essentially independent of $g_{a(\phi)f}$. 
For smaller couplings, below the lower boundary in the khaki region, energy transfer from the SM is inefficient, leading to an underproduction of dark matter.
 For the masses considered, the couplings of the mediators to SM fermions are always feeble, with $g_{a(\phi)f}$ in the range  $10^{-10}$ to $10^{-13}$ GeV$^{-1}$ while the mediators couplings to DM are weak with  $g_{a(\phi)\chi}$ in the range $10^{-3}$ to $10^{-2}$ GeV$^{-1}$.

In Fig.~\ref{fig:constraints}, the CMB exclusions are shaded in red. For $m_{\phi(a)}=250~{\rm MeV}$ (top-left panel), the dominant mediator decay channel is into $\mu^+ \mu^-$, as listed in Tab.~\ref{tab:decaywidthpseudoscalar}. The primary sources of photons and positrons in this case are muon decays and associated radiative (FSR) processes, notably $\mu^+ \to e^+ \nu_e \bar{\nu}_{\mu}$. In the bottom-left panel with mediator masses around $3~{\rm GeV}$, the dominant annihilation products are gluons and $s$ quarks, resulting in hadronic cascades that shape the photon and positron spectra. For the bottom-right panel with $m_{\phi(a)} = 30~\rm{GeV}$, annihilation predominantly proceeds via $4\tau$ or $4b$ final states, depending on the available phase space and the mediator's branching ratios to SM fermions.
The shape of the CMB exclusion varies across mediator masses. In the 30~GeV panel, the upturn at low $m_\chi$ is a clear consequence of phase space suppression near the $\chi\bar\chi \to a\phi$ kinematic threshold $m_\chi \approx m_a+m_\phi$,  the reduced phase space sharply suppresses $\langle \sigma v \rangle$, thereby requiring higher couplings to stay within the excluded region. The 250~MeV case, on the other hand, exhibits a steady slope, as the final state ($4\mu$) remains fully accessible over the entire range and suffers minimal threshold suppression. For the 3~GeV case, the bound declines more gradually without a pronounced shape feature, reflecting a mixture of softer final states (dominated by gluons) and mild threshold effects. Overall, the behaviour is consistent with an $s$-wave annihilation channel. Finally, the exclusion becomes noticeably tighter for larger DM masses across all three panels. While this might naively appear contradictory, it arises from the steep scaling of the $\chi \bar{\chi} \to s\, a$ cross section, which goes as $\sim g_{a(\phi)\chi}^4 m_\chi^2$. Although the bound on $\langle \sigma v \rangle$ weakens with increasing $m_\chi$, as expected from Eq.~\ref{eq:cmb}, the required coupling $g_{a(\phi)\chi}$ still decreases to satisfy the constraint. 

We show \texttt{MeerKAT} sensitivities in blue in Fig.~\ref{fig:constraints}. In each subfigure, the dashed blue lines give an idea of the uncertainty on these constraints, estimated by varying key astrophysical inputs: the DM halo parameters (e.g.~scale density and scale radius), the slope of the density profile, and the magnetic field strength in the cluster.
The solid blue line indicates the conservative exclusion limit, derived using the minimum scale density~\cite{Beck:2023oza}, a modified NFW profile with $\alpha = 0.5$, and a peak magnetic field strength of $B_0 = 2~\mu\rm{G}$. The dashed blue line corresponds to the median scenario, assuming a standard NFW profile, $B_0 = 8~\mu\rm{G}$, and central values for the halo parameters. 
We adopt the conservative estimate by shading the region above the solid blue line in each subfigure, indicating the excluded parameter space in $\langle \sigma v \rangle$, which, as in the case of the CMB limits, is directly translated into a constraint on $g_{a(\phi)\chi}$. This follows from the fact that, in the $g_{a\chi} = g_{\phi\chi}$ limit, DM annihilation is dominated by the $s$-wave process $\chi\bar{\chi} \to a\phi$, with the scaling relation $g_{a(\phi)\chi}\, \propto\, \langle \sigma v \rangle^{1/4}$.
The shape of the radio constraints reflects the cascade of DM annihilation products into $e^\pm$, which then emit synchrotron radiation in the galactic magnetic field. Since the $4f$ final states vary across the benchmark points (e.g., $4\mu$, $4g$, $4b$ or $4\tau$), the resulting $e^\pm$ spectra, and hence the radio flux, differ significantly. Combined with the $s$-wave scaling $\langle \sigma v \rangle\, \propto\, g_{a(\phi)\chi}^4 m_\chi^2$, this leads to nontrivial dependence of the exclusion shape on the $g_{a(\phi)\chi}$ vs.~$m_\chi$ plane.

Lastly, we show \texttt{Fermi-LAT} exclusions as green shaded regions in the $g_{a(\phi)\chi}$ vs.~$m_\chi$ plane, included for mediator masses $m_{a(\phi)} = 3$ and $30~\rm{GeV}$. The dominance of heavy quarks and gluon in the annihilation channels make \texttt{Fermi-LAT} most effective, since the leading channels are $\chi\bar{\chi} \to a\phi$, followed by $a(\phi) \to b\bar{b},\ c\bar{c},\ \tau^+\tau^-$ and $GG$, all of which are strong sources of hard-photons in the \texttt{Fermi-LAT} energy range. For the 250~MeV case, the final state is leptons, leading to a weaker resulting photon flux, and therefore, the CMB and radio constraints turn out to be stronger by several orders of magnitude.
Note that the 30~GeV \texttt{Fermi-LAT} exclusion nearly overlaps with the {\it Secluded} region, meaning that, in practice, only the 3~GeV case is independently constrained by \texttt{Fermi-LAT}.

As discussed in Sec.~\ref{sec:constraints}, BBN constrains mediator lifetimes, since late decays can distort the primordial element abundances. These lifetime limits map onto bounds on the mediator–SM couplings $g_{af}$ and $g_{\phi f}$, which govern the decay rates of $a$ and $\phi$.
To evaluate these constraints within our model, we compute the lifetimes and cosmological abundances of $a$ and $\phi$ across the parameter space shown in Figs.~\ref{fig:constraints}. 
The relevant decay widths for both mediators are listed in Tabs.~\ref{tab:decaywidthscalar} and~\ref{tab:decaywidthpseudoscalar}, from which we determine their lifetimes. Mediator abundances are again obtained by solving the Boltzmann equations~\eqref{eq:generalBoltzmann} and~\eqref{eq:EBEinT} numerically for the  combination of couplings in Figs.~\ref{fig:constraints}, until the comoving $a$ and $\phi$ number densities stay constant.

The scalar $\phi$ and pseudoscalar $a$ do not, in general, have identical cosmological abundances or decay lifetimes.  For each species we determine the minimum lifetime that would be excluded by BBN, given its individual abundance. 
A key result is that, for all couplings that yield a lifetime near the BBN sensitivity threshold, each mediator is already sufficiently abundant on its own to saturate the BBN constraint. In other words, whenever $\phi$ (or $a$) has a long enough lifetime to be constrained, its abundance is already large enough that adding the contribution from the other mediator does not change the limit.
Therefore, the two mediators do not need to be combined into a joint constraint. Their BBN exclusions can be treated independently, and at each parameter point we simply adopt whichever of the two individual exclusions is stronger. 
We will now proceed to discuss the case of each mediator mass separately.

For $m_a=m_\phi=250$ MeV the mediator particles decay almost exclusively into muons, see Tabs.~\ref{tab:decaywidthscalar} and \ref{tab:decaywidthpseudoscalar}. 
At these small masses, hadron production is kinematically forbidden, making photodissociation constraints due to electromagnetic showers most relevant. 
The muons decay almost exclusively into electrons and neutrinos ($\mu^- \rightarrow e^{-} \bar{\nu}_e \nu_\mu$). 
We apply here the constraints from Fig.~5 of~\cite{Kawasaki:2020qxm} which provides 95 $\%$ CL upper bounds for decays into electrons on $m_{a(\phi)} Y_{a(\phi)}/2$, where $Y_{a(\phi)}$ is the comoving  density of $a(\phi)$ particles after freeze-out, as a function of their lifetime $\tau_{a(\phi)}$. 
The bounds provided for masses $m_{a(\phi)}=100$ MeV and $m_{a(\phi)}=1$ GeV are interpolated to estimate constraints for $m_{a(\phi)}=250$ MeV, excluding abundant mediator particles with lifetimes above $\sim 100$ seconds. 
Although the bounds discussed above are derived for decays into electrons, they are expected to apply equally well to muon decays \cite{Protheroe:1994dt,Kawasaki:1994sc}. 
At high injection energies, the subsequent electromagnetic cascades  depend mainly on the total injected energy rather than on the nature of the primary decay products.
Consequently, as shown in~\cite{Coffey:2020oir} for the case of sub-GeV vector bosons decaying into muons, significant differences between electron and muon final states appear only for lifetimes $\tau \gtrsim 10^7~s$.  
For shorter lifetimes, the constraints remain nearly identical.\footnote{Ref.~\cite{1110.2895} suggests that constraints for muon decays could be further weakened because the additional neutrinos from the decay increase $N_{\mathrm{eff}}$; and astrophysical measurements actually favour a higher value. 
However, the most stringent constraint from~\cite{Kawasaki:2020qxm} arises from the abundance of deuterium and is mainly due to the photodissociation of D.}
For the relevant $g_{af}$ and $g_{\phi f}$ couplings, the $a$ and $\phi$ are abundant enough to be maximally constrained, with lifetimes $\tau\sim 100$ s excluded. 
These constraints are hence treated separately. The bound  from $\phi \rightarrow \mu^+\mu^-$ decays is shown in the top panel of Fig.~\ref{fig:constraints} (turquoise shaded region). 
We do not depict the corresponding constraint for the pseudoscalar decay $a \rightarrow \mu^+\mu^-$ since it is a factor of $\sim 1.4$ weaker. 

For $m_a=m_\phi=$ 3 GeV, hadronic decay channels open up, see Tabs.~\ref{tab:decaywidthscalar} and \ref{tab:decaywidthpseudoscalar}. 
The most stringent bound arises from mediator decays into gluons, which is the dominant decay channel for both the scalar and the pseudoscalar at this mass. 
To determine the  bounds shown in the bottom left of Fig.~\ref{fig:constraints}  the mediator abundances were weighted by the respective branching fractions for $\phi \rightarrow gg$ and $a \rightarrow gg$ (Tabs.~\ref{tab:decaywidthscalar} and \ref{tab:decaywidthpseudoscalar}). 
Using Fig.~12 of~\cite{Kawasaki:2017bqm}, we find that these bounds constrain mediator-fermion couplings as large as $\sim 3\times 10^{-11}$ GeV$^{-1}$, corresponding to lifetimes of $\sim 5\times 10^{-2}$ s. 
Mediator abundances in this parameter region are sufficiently large to independently constrain the lifetime. The constraint in Fig.~\ref{fig:constraints} comes from decays $\phi \rightarrow gg$ which is again stronger than the one from $a \rightarrow gg$ and excludes a significant fraction of the DFO region. 
Constraints from electromagnetic decays are orders of magnitude weaker.

 For $m_{a}=m_\phi=30$ GeV lifetimes of the order of $\sim 5\times 10^{-2}$ s corresponding to values of $g_{a(\phi)f}\sim 5\times 10^{-13}$ GeV$^{-1}$ can be excluded, coming from constraints on the scalar decaying into bottom quarks, $\phi\rightarrow b\bar{b}$. In the upper half of the DFO region the mediator particles have lifetimes too short to be constrained by BBN. 

 Note that in Fig.~\ref{fig:constraints} the BBN exclusion lines in the $g_{a(\phi)\chi}$–$m_\chi$ plane are evaluated for a fixed value of $g_{a(\phi)f}$, which controls the decay rate and hence the lifetime relevant for BBN. Since the BBN bounds depend primarily on $g_{a(\phi)f}$ and only weakly on $g_{a(\phi)\chi}$, each exclusion curve should be interpreted as corresponding to a particular choice of $g_{a(\phi)f}$, effectively tracing a single lifetime contour. 

\section{Comparison with secluded freeze-out}
\label{sec:secluded}

In the previous sections, we have argued that
 the DFO mechanism permits some cosmological and astrophysical constraints to be relaxed. To show this in the framework of the model we have considered so far,  we  compare the decoupled  FO with the secluded FO~\cite{Pospelov:2007mp,NFortes:2022dkj} scenario explicitly. In the latter,
DM annihilation into mediators governs DM formation, and the couplings of the mediators to the SM are required to be large enough to bring the mediators into thermal equilibrium with the SM bath. Typically couplings $\ge 10^{-6}$ are sufficient,  thus for definiteness we will fix   $g_{aq}=g_{\phi q}=10^{-6}$  and we furthermore always assume equal couplings to quark and leptons.  

Under the assumption of thermal equilibrium, the DM relic density, the CMB constraint as well as the radio constraint from \texttt{MeerKAT} only depend on  $g_{a\chi}g_{\phi \chi}$. 
Direct detection (DD)  of DM scattering on nucleons on the other hand is driven  primarily by the exchange of the scalar $\phi$ and thus depends on the square of the product of the couplings $g_{\phi \chi} g_{\phi q}$, here we neglect possible mixing of the scalar with the SM Higgs.  The results from DD strongly constrain the case where the mediator is light since the spin-independent elastic cross-section goes as $1/m_\phi^4$. For the case of a light mediator, $m_a=m_\phi=250~\rm{MeV}$, \texttt{PandaX}~\cite{PandaX-4T:2021bab} (\texttt{LZ}~\cite{LZ:2024zvo}) excludes $g_{\phi \chi} g_{\phi q} \approx 10^{-11}(10^{-12})$, while a coupling $g_{\phi \chi} \approx 10^{-3}$ is required to satisfy the relic density constraint.\footnote{Note that
it would be possible to escape DD limits for DM masses below 3.5 GeV, but  this region is completely ruled out by CMB measurements.} We therefore conclude that for a light mediator mass,  $m_a=m_\phi=250~\rm{MeV}$,  the secluded scenario is  ruled out. 
 
 \begin{figure}
    \centering
    \includegraphics[width=0.7\linewidth]{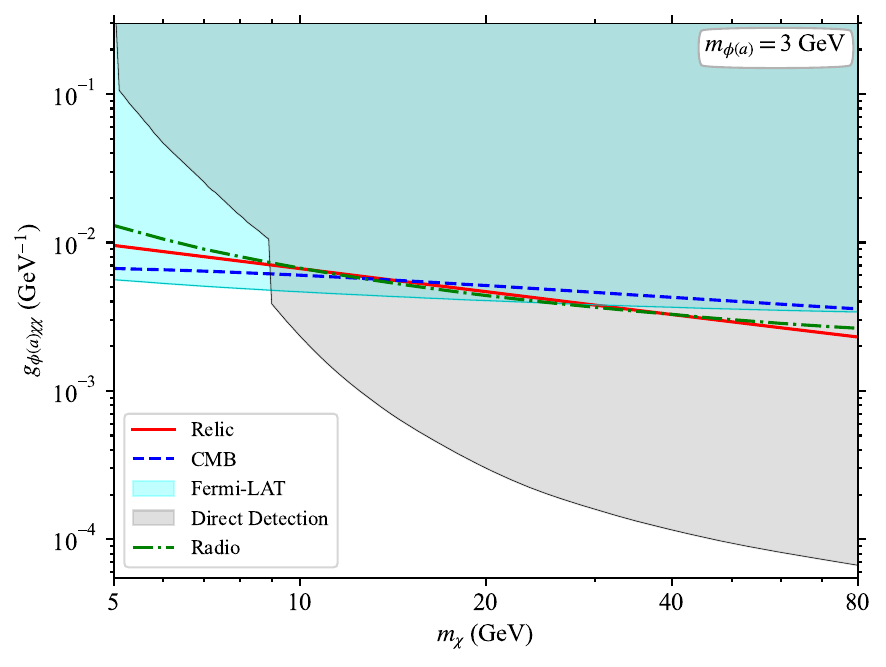}
      \caption{Constraints on the parameter space of the secluded FO scenario with  $m_a=m_\phi=3\ {\rm GeV}$ and $g_{aq}=g_{\phi q}=10^{-6}$.  The direct detection limit uses a combination of \texttt{PandaX}~\cite{PandaX-4T:2021bab} and \text{LZ~2024}~\cite{LZ:2024zvo}. }
         \label{fig:secluded}
\end{figure}
 
As the mass of the mediator increases, the DD constraint relaxes significantly, nevertheless, as seen in 
Fig.~\ref{fig:secluded} which shows  the contour for $\Omega h^2=0.12$   in the $g_{a(\phi)\chi} - m_\chi$ plane for $m_a= m_\phi=3~{\rm GeV}$, the region satisfying the relic density constraint is excluded by DD for masses above 9 GeV. 
Moreover the CMB limit excludes all DM masses below about 15 GeV while the 
conservative bound from radio data from  \texttt{MeerKAT} constrains the DM mass in the range  15-40 GeV.  Finally, the \texttt{Fermi-LAT} limit on photons from dSPhs. galaxies rules out DM below about 30~GeV, leaving again no allowed space for the secluded scenario.
The DD constraint can be further  relaxed when considering a heavier mediator, we found that for $m_a= m_\phi=30~{\rm GeV}$ and $g_{aq}=g_{\phi q}=10^{-6}$, the entire relic density favoured region evades DD constraints, thus allowing secluded FO for $m_\chi> 30~{\rm GeV}$. 
Note, however, that this relies on our assumption for the  value of the mediator couplings to quarks, increasing $g_{aq}=g_{\phi q}=10^{-4}$ would lead to the exclusion of the scenario by \texttt{LZ}.

\section{Discussion and conclusion}
\label{sec:discussion}
  In this paper, we have focused on probing decoupled freeze-out through DM indirect detection, which is typically challenging for non-thermal dark matter scenarios. By considering a setup with dominant $s$-wave annihilation, we identify a highly testable parameter region within the reach of CMB, Fermi-LAT and radio observations, with indirect detection bounding the parameter space from above and BBN from below, leaving a viable region. We show in a simple model where we add a dark matter particle and two mediators to the SM,  DM formation via decoupled FO mechanism can be achieved for mediator and dark matter masses of the order of the electroweak scale or below. While strong constraints arise from indirect detection and from the CMB, owing to the DM annihilation process being $s$-wave, as well as from BBN, direct detection plays no role because the couplings of the mediators to quarks is feeble. This is in contrast with the secluded FO scenario where the larger couplings of the mediators to quarks lead to important constraints from direct detection. Thus we conclude  that DFO scenarios are more widely viable than the secluded FO scenario.

Note however that all our results rely on the assumption that $g_{a\chi}=g_{\phi\chi}$. 
As we tested numerically and can intuitively understand, this case  is 
quite generic. The region that yields the correct relic density is governed by the interaction $\chi \bar \chi \leftrightarrow a \phi$, which is $s$-wave  and inversely proportional to the coupling combination $(g_{a\chi}\,g_{\phi\chi})^2$. Both CMB and indirect detection constraints depend on the same combination of couplings. 
Thus our results -- for the relic density, CMB and indirect detection  constraints --- remain valid for different ratios $r=g_{a\chi}/g_{\phi\chi}$  as long as the product of the coupling remains constant and the two couplings are of the same order. Changing the ratio $r>1$ will lead to a shift upward of the  allowed region in the  $g_{a\chi}-m_\chi$ plane by a factor of approximately $\sim \sqrt{r}$. Since the bounds from BBN are independent of $g_{a\chi}$ and $g_{\phi\chi}$ they will also shift upward and so will the bounds from CMB and indirect detection. However when  there is a large splitting between $g_{a\chi}$ and $g_{\phi\chi}$, we expect a different behaviour.

On the other hand, if $r \gg 1$ the dominant process for DM annihilation will switch to $\chi\bar \chi \to aa$ which is a $p$-wave process. The relic density constraint will require a larger value of the coupling $g_{a\chi}$, while the CMB and indirect detection constraints will  weaken  until at some point they will decouple when the $p$-wave process dominates. For example, in the case $m_\chi=10~{\rm GeV},\, m_a=3~{\rm GeV}$ one  escapes the \texttt{Fermi-LAT} constraint when $r\ge 10$ and  
 the CMB constraint is relaxed by a factor 2 for $r>14$. For these values of couplings the relic density receives roughly equal contributions from $aa$ and $a\phi$ processes.
The same phenomenon is observed in the limit $r \ll 1$ where  the main process for DM annihilation will switch to $\chi \bar \chi \to \phi\phi$, again when the $p$-wave contribution becomes dominant both the \texttt{Fermi-LAT} and CMB constraint relax significantly. 

  The DFO mechanism will also be probed by future dark matter searches. Among the lab-based experiments, the \texttt{LZXD} direct detection search~\cite{Aalbers:2022dzr} can  probe some of the allowed region when the mediator is light. For example, for $m_\phi=250~{\rm MeV}$, the couplings $g_{\phi q}= 3.1\times 10^{-11} (10 ^{-10})$ are within the reach of \texttt{LZXD} for $m_\chi=30 (80)$ GeV.
The future sky-surveys such as \texttt{CMB-S4} mission is expected to increase the precision on DM annihilation from measurements of CMB anisotropies by a factor 2-3~\cite{Dvorkin:2022bsc}. This will only mildly improve the constraint on the DM couplings in our model since $\langle \sigma v \rangle\, \propto\, (g_{a\chi}\, g_{\phi\chi})^2$.
Future indirect detectors that are sensitive to low-energy photons such as \texttt{e-ASTROGAM}~\cite{e-ASTROGAM:2017pxr} or \texttt{MAST}~\cite{Dzhatdoev:2019kay} can probe dark matter below 10 GeV. However, Ref.~\cite{ODonnell:2024aaw} showed that for DM annihilation into muons --- as in the case for the light mediators--  future detectors can probe mainly the region compatible with the current cosmological bounds. Also, future radio telescopes such as \texttt{SKA}~\cite{Wang:2023sxr} are expected to provide robust limits on the same parameter space with sensitivities extending even into the sub-GeV regime. Altogether, these developments will continue to narrow down the viable parameter space, offering complementary coverage across detection strategies.

While our analysis assumes negligible direct couplings between the mediators, a richer scalar/pseudoscalar sector with more elaborate interactions could give rise to a broader range of thermal histories. For instance, if the mediators are strongly coupled to one another, they could share a common dark temperature regardless of whether the DM itself is thermalised. In such scenarios, the mediators effectively form their own thermal bath, and depending on the strength of their coupling to the DM, the latter may undergo either freeze-in or freeze-out. Exploring such an extended mediator structure could reveal a richer phase diagram of DM production mechanisms, and would be an interesting direction for future work.\\

\noindent
{\bf Acknowledgements}

\noindent
SC acknowledges support from the UKRI Future Leader Fellowship DarkMAP (Ref. no. MR/Y034112/1). RI was supported by the SERB-TARE research grant (TAR/2020/000448) of Department of Science and Technology, Govt. of India.
SM was supported through the European Union (ERC,
QFT.zip project, Grant Agreement no.~101040260). 
SC and GB acknowledge support from Institut Pascal and the P2I axis of the Graduate School of Physics during the Paris-Saclay Astroparticle Symposium 2025, as well as from the CNRS IRP UCMN, where the final stages of this work were completed. AB, SC and SM thank the organisers of the axions++ workshop where the collaboration began.

\appendix

\section{Decay Widths}\label{app:decaywidths}

According to \eqref{eq:LagDS-ferm2}, the mediators ($\phi, a$) in our model can decay into the SM fermions depending on the respective masses of the particles involved. Although our model has no direct coupling between the dark sector particles with the SM gauge bosons, the scalar-fermion interactions \eqref{eq:LagDS-ferm2} might give rise to such couplings via loops. We have taken into account such couplings through the effective interactions \eqref{eq:LagDS-gluon}. As a result, we get  the following partial decay widths for the scalar mediator $\phi$ to the SM particles, and show in Tab.~\ref{tab:decaywidthscalar}  the branching fractions for some representative values of $m_\phi$.

\begin{align}
    \Gamma(\phi\to f\bar f)
    &=
    N_f \frac{g^2_{\phi f}}{8\pi} \frac{m^2_f}{m^2_\phi} \big[m^2_\phi - 4m^2_f\big]^{3/2};
    \\
    \Gamma(\phi \to gg)
    &=
    \frac{m^3_\phi}{32\pi^3} \alpha^2_S(m^2_\phi) \Big|\sum_q g_{\phi q} F^q_G\Big|^2.
\end{align}

\begin{table}[!ht]
    \centering
    \begin{tabular}{c|c|c|c|c}
        \hline
        Channel & \makecell{Scalar mass \\ {[GeV]}} & \makecell{Mass of decaying \\ particles [GeV]} & \makecell{Decay width \\ {[GeV]}} & \makecell{Branching \\ Fraction [\%]} \\
        \hline\hline
        $\mu\bar\mu$ & \multirow{5}*{30} & $105.7 \times 10^{-3}$ & $1.334 \times 10^{-2}$ & $0.02$ \\
        \hhline{-~---}
        $\tau\bar\tau$ & & $1.777$ & $3.690$ & $5.66$ \\
        \hhline{-~---}
        $c\bar c$ & & $1.270$ & $5.714$ & $8.76$ \\
        \hhline{-~---}
        $b\bar b$ & & $4.180$ & $55.42$ & $84.99$ \\
        \hhline{-~---}
        $gg$ & & $0$ & $0.342$ & $0.52$ \\
        \hline\hline
        $\mu\bar\mu$ & \multirow{3}*{3} & $105.7 \times 10^{-3}$ & $1.324 \times 10^{-3}$ & $7.19$ \\
        \hhline{-~---}
        $s\bar s$ & & $93.4 \times 10^{-3}$ & $3.106 \times 10^{-3}$ & $16.88$ \\
        \hhline{-~---}
        $gg$ & & $0$ & $1.397 \times 10^{-2}$ & $75.93$ \\
        \hline\hline
        $e\bar e$ & \multirow{2}*{$250\times10^{-3}$} & $0.511 \times 10^{-3}$ & $2.597 \times 10^{-9}$ & $0.02$ \\
        \hhline{-~---}
        $\mu\bar\mu$ & & $105.7 \times 10^{-3}$ & $1.691 \times 10^{-5}$ & $99.98$ \\
        \hline
    \end{tabular}
    \caption{Decay channels of the scalar mediator $\phi$ for $g_{\phi f}=$1 GeV$^{-1}$}
    \label{tab:decaywidthscalar}
\end{table}

In a similar vein, we calculate the decay widths for the pseudoscalar mediator $a$ to the SM particles and show in Tab.~\ref{tab:decaywidthpseudoscalar} an estimate of the branching fractions for some representative values of $m_a$.
\begin{align}
    \Gamma(a\to f\bar f)
    &=
    N_f \frac{g^2_{af}}{8\pi} m^2_f \big[m^2_a - 4m^2_f\big]^{1/2};
    \\
    \Gamma(s \to gg)
    &=
    \frac{m^3_a}{32\pi^3} \alpha^2_S(m^2_a) \Big|\sum_q g_{aq} \widetilde{F}^q_G\Big|^2.
\end{align}

\begin{table}[!ht]
    \centering
    \begin{tabular}{c|c|c|c|c}
        \hline
        Channel & \makecell{Pseudoscalar \\ mass [GeV]} & \makecell{Mass of decaying \\ particles [GeV]} & \makecell{Decay width \\ {[GeV]}} & \makecell{Branching \\ Fraction [\%]} \\
        \hline\hline
        $\mu\bar\mu$ & \multirow{5}*{30} & $105.7 \times 10^{-3}$ & $1.334 \times 10^{-2}$ & $0.02$ \\
        \hhline{-~---}
        $\tau\bar\tau$ & & $1.777$ & $3.743$ & $5.33$ \\
        \hhline{-~---}
        $c\bar c$ & & $1.270$ & $5.755$ & $8.20$ \\
        \hhline{-~---}
        $b\bar b$ & & $4.180$ & $60.09$ & $85.65$ \\
        \hhline{-~---}
        $gg$ & & $0$ & $0.555$ & $0.79$ \\
        \hline\hline
        $\mu\bar\mu$ & \multirow{3}*{3} & $105.7 \times 10^{-3}$ & $1.330 \times 10^{-3}$ & $3.65$ \\
        \hhline{-~---}
        $s\bar s$ & & $93.4 \times 10^{-3}$ & $3.118 \times 10^{-3}$ & $8.56$ \\
        \hhline{-~---}
        $gg$ & & $0$ & $3.196 \times 10^{-2}$ & $87.78$ \\
        \hline\hline
        $e\bar e$ & \multirow{2}*{$250\times10^{-3}$} & $0.511 \times 10^{-3}$ & $2.597 \times 10^{-9}$ & $0$ \\
        \hhline{-~---}
        $\mu\bar\mu$ & & $105.7 \times 10^{-3}$ & $5.933 \times 10^{-5}$ & $100$ \\
        \hline
    \end{tabular}
     \caption{Decay channels of the pseudoscalar mediator $a$ for $g_{a f}=$1 GeV$^{-1}$}
     \label{tab:decaywidthpseudoscalar}
\end{table}

\bibliography{biblio}

@inproceedings{Jonas:2018Jr,
  author = "Jonas, Justin",
  title = "{The MeerKAT Radio Telescope}",
  doi = "10.22323/1.277.0001",
  booktitle = "Proceedings of MeerKAT Science: On the Pathway to the SKA {\textemdash} PoS(MeerKAT2016)",
  year = 2018,
  volume = "277",
  pages = "001"
}

@article{McDonald:2001vt,
    author = "McDonald, John",
    title = "{Thermally generated gauge singlet scalars as selfinteracting dark matter}",
    eprint = "hep-ph/0106249",
    archivePrefix = "arXiv",
    doi = "10.1103/PhysRevLett.88.091304",
    journal = "Phys. Rev. Lett.",
    volume = "88",
    pages = "091304",
    year = "2002"
}

@article{Hall:2009bx,
    author = "Hall, Lawrence J. and Jedamzik, Karsten and March-Russell, John and West, Stephen M.",
    title = "{Freeze-In Production of FIMP Dark Matter}",
    eprint = "0911.1120",
    archivePrefix = "arXiv",
    primaryClass = "hep-ph",
    reportNumber = "OUTP-09-18-P, UCB-PTH-09-32",
    doi = "10.1007/JHEP03(2010)080",
    journal = "JHEP",
    volume = "03",
    pages = "080",
    year = "2010"
}

@article{Bernal:2017kxu,
    author = "Bernal, Nicol{\'a}s and Heikinheimo, Matti and Tenkanen, Tommi and Tuominen, Kimmo and Vaskonen, Ville",
    title = "{The Dawn of FIMP Dark Matter: A Review of Models and Constraints}",
    eprint = "1706.07442",
    archivePrefix = "arXiv",
    primaryClass = "hep-ph",
    reportNumber = "PI-UAN-2017-602FT, HIP-2017-08-TH, PI-UAN--2017--602FT, HIP--2017--08-TH",
    doi = "10.1142/S0217751X1730023X",
    journal = "Int. J. Mod. Phys. A",
    volume = "32",
    number = "27",
    pages = "1730023",
    year = "2017"
}

@article{Feng:2003uy,
    author = "Feng, Jonathan L. and Rajaraman, Arvind and Takayama, Fumihiro",
    title = "{SuperWIMP dark matter signals from the early universe}",
    eprint = "hep-ph/0306024",
    archivePrefix = "arXiv",
    reportNumber = "UCI-TR-2003-19",
    doi = "10.1103/PhysRevD.68.063504",
    journal = "Phys. Rev. D",
    volume = "68",
    pages = "063504",
    year = "2003"
}

@article{Hambye:2019dwd,
    author = "Hambye, Thomas and Tytgat, Michel H. G. and Vandecasteele, J{\'e}r{\^o}me and Vanderheyden, Laurent",
    title = "{Dark matter from dark photons: a taxonomy of dark matter production}",
    eprint = "1908.09864",
    archivePrefix = "arXiv",
    primaryClass = "hep-ph",
    reportNumber = "ULB-TH/19-07",
    doi = "10.1103/PhysRevD.100.095018",
    journal = "Phys. Rev. D",
    volume = "100",
    number = "9",
    pages = "095018",
    year = "2019"
}

@article{Coy:2021ann,
    author = "Coy, Rupert and Hambye, Thomas and Tytgat, Michel H. G. and Vanderheyden, Laurent",
    title = "{Domain of thermal dark matter candidates}",
    eprint = "2105.01263",
    archivePrefix = "arXiv",
    primaryClass = "hep-ph",
    reportNumber = "ULB-TH/21-05",
    doi = "10.1103/PhysRevD.104.055021",
    journal = "Phys. Rev. D",
    volume = "104",
    number = "5",
    pages = "055021",
    year = "2021"
}

@article{Hambye:2020lvy,
    author = "Hambye, Thomas and Lucca, Matteo and Vanderheyden, Laurent",
    title = "{Dark matter as a heavy thermal hot relic}",
    eprint = "2003.04936",
    archivePrefix = "arXiv",
    primaryClass = "hep-ph",
    reportNumber = "ULB-TH/20-02, ULB-TH/20-02",
    doi = "10.1016/j.physletb.2020.135553",
    journal = "Phys. Lett. B",
    volume = "807",
    pages = "135553",
    year = "2020"
}

@article{Belanger:2020npe,
    author = "B{\'e}langer, Genevi{\`e}ve and Delaunay, C{\'e}dric and Pukhov, Alexander and Zaldivar, Bryan",
    title = "{Dark matter abundance from the sequential freeze-in mechanism}",
    eprint = "2005.06294",
    archivePrefix = "arXiv",
    primaryClass = "hep-ph",
    reportNumber = "LAPTH-021/20, IFT-UAM/CSIC-20-65",
    doi = "10.1103/PhysRevD.102.035017",
    journal = "Phys. Rev. D",
    volume = "102",
    number = "3",
    pages = "035017",
    year = "2020"
}

@article{Chu:2011be,
    author = "Chu, Xiaoyong and Hambye, Thomas and Tytgat, Michel H. G.",
    title = "{The Four Basic Ways of Creating Dark Matter Through a Portal}",
    eprint = "1112.0493",
    archivePrefix = "arXiv",
    primaryClass = "hep-ph",
    reportNumber = "ULB-TH-11-26",
    doi = "10.1088/1475-7516/2012/05/034",
    journal = "JCAP",
    volume = "05",
    pages = "034",
    year = "2012"
}

@article{Binder:2017rgn,
    author = "Binder, Tobias and Bringmann, Torsten and Gustafsson, Michael and Hryczuk, Andrzej",
    title = "{Early kinetic decoupling of dark matter: when the standard way of calculating the thermal relic density fails}",
    eprint = "1706.07433",
    archivePrefix = "arXiv",
    primaryClass = "astro-ph.CO",
    doi = "10.1103/PhysRevD.96.115010",
    journal = "Phys. Rev. D",
    volume = "96",
    number = "11",
    pages = "115010",
    year = "2017",
    note = "[Erratum: Phys.Rev.D 101, 099901 (2020)]"
}

@article{Binder:2021bmg,
    author = "Binder, Tobias and Bringmann, Torsten and Gustafsson, Michael and Hryczuk, Andrzej",
    title = "{Dark matter relic abundance beyond kinetic equilibrium}",
    eprint = "2103.01944",
    archivePrefix = "arXiv",
    primaryClass = "hep-ph",
    doi = "10.1140/epjc/s10052-021-09357-5",
    journal = "Eur. Phys. J. C",
    volume = "81",
    pages = "577",
    year = "2021"
}

@article{Belanger:2024bro,
    author = "B{\'e}langer, Genevieve and Chakraborti, Sreemanti and G{\'e}nolini, Yoann and Salati, Pierre",
    title = "{GeV-scale dark matter with p-wave Breit-Wigner enhanced annihilation}",
    eprint = "2401.02513",
    archivePrefix = "arXiv",
    primaryClass = "hep-ph",
    reportNumber = "LAPTH-002/24, IPPP/23/85",
    doi = "10.1103/PhysRevD.110.023039",
    journal = "Phys. Rev. D",
    volume = "110",
    number = "2",
    pages = "023039",
    year = "2024"
}

@article{Belanger:2025kce,
    author = "B{\'e}langer, Genevi{\`e}ve and Chakraborti, Sreemanti and Delaunay, C{\'e}dric and Jomain, Margaux",
    title = "{Rekindling s-Wave Dark Matter Annihilation Below 10~GeV with Breit-Wigner Effects}",
    eprint = "2503.08897",
    archivePrefix = "arXiv",
    primaryClass = "hep-ph",
    reportNumber = "IPPP/25/14, LAPTH-009/25",
    month = "3",
    year = "2025"
}

@article{Cirelli:2024ssz,
    author = "Cirelli, Marco and Strumia, Alessandro and Zupan, Jure",
    title = "{Dark Matter}",
    eprint = "2406.01705",
    archivePrefix = "arXiv",
    primaryClass = "hep-ph",
    month = "6",
    year = "2024"
}

@article{Belanger:2011ww,
    author = "Belanger, Genevieve and Park, Jong-Chul",
    title = "{Assisted freeze-out}",
    eprint = "1112.4491",
    archivePrefix = "arXiv",
    primaryClass = "hep-ph",
    doi = "10.1088/1475-7516/2012/03/038",
    journal = "JCAP",
    volume = "03",
    pages = "038",
    year = "2012"
}

@article{Pospelov:2007mp,
    author = "Pospelov, Maxim and Ritz, Adam and Voloshin, Mikhail B.",
    title = "{Secluded WIMP Dark Matter}",
    eprint = "0711.4866",
    archivePrefix = "arXiv",
    primaryClass = "hep-ph",
    doi = "10.1016/j.physletb.2008.02.052",
    journal = "Phys. Lett. B",
    volume = "662",
    pages = "53--61",
    year = "2008"
}

@article{DAgnolo:2015ujb,
    author = "D'Agnolo, Raffaele Tito and Ruderman, Joshua T.",
    title = "{Light Dark Matter from Forbidden Channels}",
    eprint = "1505.07107",
    archivePrefix = "arXiv",
    primaryClass = "hep-ph",
    doi = "10.1103/PhysRevLett.115.061301",
    journal = "Phys. Rev. Lett.",
    volume = "115",
    number = "6",
    pages = "061301",
    year = "2015"
}

@article{Aalbers:2022dzr,
    author = "Aalbers, J. and others",
    title = "{A next-generation liquid xenon observatory for dark matter and neutrino physics}",
    eprint = "2203.02309",
    archivePrefix = "arXiv",
    primaryClass = "physics.ins-det",
    reportNumber = "INT-PUB-22-003, FERMILAB-PUB-22-112-PPD-QIS-T",
    doi = "10.1088/1361-6471/ac841a",
    journal = "J. Phys. G",
    volume = "50",
    number = "1",
    pages = "013001",
    year = "2023"
}

@article{DAgnolo:2020mpt,
    author = "D'Agnolo, Raffaele Tito and Liu, Di and Ruderman, Joshua T. and Wang, Po-Jen",
    title = "{Forbidden dark matter annihilations into Standard Model particles}",
    eprint = "2012.11766",
    archivePrefix = "arXiv",
    primaryClass = "hep-ph",
    doi = "10.1007/JHEP06(2021)103",
    journal = "JHEP",
    volume = "06",
    pages = "103",
    year = "2021"
}

@article{NFortes:2022dkj,
    author = "N. Fortes, Guilherme and S. Queiroz, Farinaldo and Siqueira, Clarissa and Viana, Aion",
    title = "{Present and future constraints on secluded dark matter in the Galactic Halo with TeV Gamma-ray observatories}",
    eprint = "2212.05075",
    archivePrefix = "arXiv",
    primaryClass = "hep-ph",
    doi = "10.1088/1475-7516/2023/07/043",
    journal = "JCAP",
    volume = "07",
    pages = "043",
    year = "2023"
}

@article{Garny:2017rxs,
    author = {Garny, Mathias and Heisig, Jan and L{\"u}lf, Benedikt and Vogl, Stefan},
    title = "{Coannihilation without chemical equilibrium}",
    eprint = "1705.09292",
    archivePrefix = "arXiv",
    primaryClass = "hep-ph",
    reportNumber = "TUM-HEP-1085-17, TTK-17-18",
    doi = "10.1103/PhysRevD.96.103521",
    journal = "Phys. Rev. D",
    volume = "96",
    number = "10",
    pages = "103521",
    year = "2017"
}

@article{DAgnolo:2017dbv,
    author = "D'Agnolo, Raffaele Tito and Pappadopulo, Duccio and Ruderman, Joshua T.",
    title = "{Fourth Exception in the Calculation of Relic Abundances}",
    eprint = "1705.08450",
    archivePrefix = "arXiv",
    primaryClass = "hep-ph",
    doi = "10.1103/PhysRevLett.119.061102",
    journal = "Phys. Rev. Lett.",
    volume = "119",
    number = "6",
    pages = "061102",
    year = "2017"
}

@article{Brummer:2019inq,
    author = {Br{\"u}mmer, F.},
    title = "{Coscattering in next-to-minimal dark matter and split supersymmetry}",
    eprint = "1910.01549",
    archivePrefix = "arXiv",
    primaryClass = "hep-ph",
    doi = "10.1007/JHEP01(2020)113",
    journal = "JHEP",
    volume = "01",
    pages = "113",
    year = "2020"
}

@article{Alguero:2022inz,
    author = "Alguero, Gael and Belanger, Genevieve and Kraml, Sabine and Pukhov, Alexander",
    title = "{Co-scattering in micrOMEGAs: A case study for the singlet-triplet dark matter model}",
    eprint = "2207.10536",
    archivePrefix = "arXiv",
    primaryClass = "hep-ph",
    doi = "10.21468/SciPostPhys.13.6.124",
    journal = "SciPost Phys.",
    volume = "13",
    pages = "124",
    year = "2022"
}

@article{Alguero:2023zol,
    author = "Alguero, G. and Belanger, G. and Boudjema, F. and Chakraborti, S. and Goudelis, A. and Kraml, S. and Mjallal, A. and Pukhov, A.",
    title = "{micrOMEGAs 6.0: N-component dark matter}",
    eprint = "2312.14894",
    archivePrefix = "arXiv",
    primaryClass = "hep-ph",
    doi = "10.1016/j.cpc.2024.109133",
    journal = "Comput. Phys. Commun.",
    volume = "299",
    pages = "109133",
    year = "2024"
}

@article{Bergstrom:1997fj,
    author = "Bergstrom, Lars and Ullio, Piero and Buckley, James H.",
    title = "{Observability of gamma-rays from dark matter neutralino annihilations in the Milky Way halo}",
    eprint = "astro-ph/9712318",
    archivePrefix = "arXiv",
    doi = "10.1016/S0927-6505(98)00015-2",
    journal = "Astropart. Phys.",
    volume = "9",
    pages = "137--162",
    year = "1998"
}

@article{Alvarez:2020cmw,
    author = "Alvarez, Alexandre and Calore, Francesca and Genina, Anna and Read, Justin and Serpico, Pasquale Dario and Zaldivar, Bryan",
    title = "{Dark matter constraints from dwarf galaxies with data-driven J-factors}",
    eprint = "2002.01229",
    archivePrefix = "arXiv",
    primaryClass = "astro-ph.HE",
    reportNumber = "LAPTH-002/20, IFT-UAM/CSIC-20-15",
    doi = "10.1088/1475-7516/2020/09/004",
    journal = "JCAP",
    volume = "09",
    pages = "004",
    year = "2020"
}

@article{Calore:2018sdx,
    author = "Calore, Francesca and Serpico, Pasquale D. and Zaldivar, Bryan",
    title = "{Dark matter constraints from dwarf galaxies: a data-driven analysis}",
    eprint = "1803.05508",
    archivePrefix = "arXiv",
    primaryClass = "astro-ph.HE",
    doi = "10.1088/1475-7516/2018/10/029",
    journal = "JCAP",
    volume = "10",
    pages = "029",
    year = "2018"
}

@misc{calore_2021_5592836,
  author       = {Calore, Francesca and
                  Zaldívar, Bryan and
                  Serpico, Pasquale and
                  Eckner, Christopher},
  title        = {Dark matter constraints from dwarf galaxies: a
                   data-driven LAT analysis
                  },
  month        = oct,
  year         = "2021",
  publisher    = {Zenodo},
  version      = {v1.0},
  doi          = {10.5281/zenodo.5592836},
  url          = {https://doi.org/10.5281/zenodo.5592836},
}

@inproceedings{Dvorkin:2022bsc,
    author = "Dvorkin, Cora and others",
    title = "{Dark Matter Physics from the CMB-S4 Experiment}",
    booktitle = "{Snowmass 2021}",
    eprint = "2203.07064",
    archivePrefix = "arXiv",
    primaryClass = "hep-ph",
    month = "3",
    year = "2022"
}

@article{e-ASTROGAM:2017pxr,
    author = "Tavani, M. and others",
    editor = "De Angelis, A. and Tatischeff, V. and Grenier, I. A. and McEnery, J. and Mallamaci, M.",
    collaboration = "e-ASTROGAM",
    title = "{Science with e-ASTROGAM: A space mission for MeV{\textendash}GeV gamma-ray astrophysics}",
    eprint = "1711.01265",
    archivePrefix = "arXiv",
    primaryClass = "astro-ph.HE",
    doi = "10.1016/j.jheap.2018.07.001",
    journal = "JHEAp",
    volume = "19",
    pages = "1--106",
    year = "2018"
}

@article{Dzhatdoev:2019kay,
    author = "Dzhatdoev, Timur and Podlesnyi, Egor",
    title = "{Massive Argon Space Telescope (MAST): A concept of heavy time projection chamber for $\gamma$-ray astronomy in the 100 MeV{\textendash}1 TeV energy range}",
    eprint = "1902.01491",
    archivePrefix = "arXiv",
    primaryClass = "astro-ph.HE",
    doi = "10.1016/j.astropartphys.2019.04.004",
    journal = "Astropart. Phys.",
    volume = "112",
    pages = "1--7",
    year = "2019"
}

@article{Cheung:2010gj,
    author = "Cheung, Clifford and Elor, Gilly and Hall, Lawrence J. and Kumar, Piyush",
    title = "{Origins of Hidden Sector Dark Matter I: Cosmology}",
    eprint = "1010.0022",
    archivePrefix = "arXiv",
    primaryClass = "hep-ph",
    reportNumber = "UCB-PTH-10-17",
    doi = "10.1007/JHEP03(2011)042",
    journal = "JHEP",
    volume = "03",
    pages = "042",
    year = "2011"
}

@article{ODonnell:2024aaw,
    author = "O'Donnell, Kayla E. and Slatyer, Tracy R.",
    title = "{Constraints on dark matter with future MeV gamma-ray telescopes}",
    eprint = "2411.00087",
    archivePrefix = "arXiv",
    primaryClass = "hep-ph",
    reportNumber = "MIT-CTP/5792",
    doi = "10.1103/PhysRevD.111.083037",
    journal = "Phys. Rev. D",
    volume = "111",
    number = "8",
    pages = "083037",
    year = "2025"
}

@article{Zurek:2008qg,
    author = "Zurek, Kathryn M.",
    title = "{Multi-Component Dark Matter}",
    eprint = "0811.4429",
    archivePrefix = "arXiv",
    primaryClass = "hep-ph",
    reportNumber = "FERMILAB-PUB-08-542-A",
    doi = "10.1103/PhysRevD.79.115002",
    journal = "Phys. Rev. D",
    volume = "79",
    pages = "115002",
    year = "2009"
}

@article{Liu:2011aa,
    author = "Liu, Ze-Peng and Wu, Yue-Liang and Zhou, Yu-Feng",
    title = "{Enhancement of dark matter relic density from the late time dark matter conversions}",
    eprint = "1101.4148",
    archivePrefix = "arXiv",
    primaryClass = "hep-ph",
    doi = "10.1140/epjc/s10052-011-1749-4",
    journal = "Eur. Phys. J. C",
    volume = "71",
    pages = "1749",
    year = "2011"
}

@article{Belanger:2012vp,
    author = "Belanger, Genevieve and Kannike, Kristjan and Pukhov, Alexander and Raidal, Martti",
    title = "{Impact of semi-annihilations on dark matter phenomenology - an example of $Z_N$ symmetric scalar dark matter}",
    eprint = "1202.2962",
    archivePrefix = "arXiv",
    primaryClass = "hep-ph",
    doi = "10.1088/1475-7516/2012/04/010",
    journal = "JCAP",
    volume = "04",
    pages = "010",
    year = "2012"
}

@article{Esch:2014jpa,
    author = "Esch, Sonja and Klasen, Michael and Yaguna, Carlos E.",
    title = "{A minimal model for two-component dark matter}",
    eprint = "1406.0617",
    archivePrefix = "arXiv",
    primaryClass = "hep-ph",
    reportNumber = "MS-TP-14-22",
    doi = "10.1007/JHEP09(2014)108",
    journal = "JHEP",
    volume = "09",
    pages = "108",
    year = "2014"
}

@article{Arcadi:2016kmk,
    author = "Arcadi, Giorgio and Gross, Christian and Lebedev, Oleg and Mambrini, Yann and Pokorski, Stefan and Toma, Takashi",
    title = "{Multicomponent Dark Matter from Gauge Symmetry}",
    eprint = "1611.00365",
    archivePrefix = "arXiv",
    primaryClass = "hep-ph",
    reportNumber = "LPT-ORSAY-16-63, HIP-2016-31-TH",
    doi = "10.1007/JHEP12(2016)081",
    journal = "JHEP",
    volume = "12",
    pages = "081",
    year = "2016"
}

@article{Feng:2008mu,
    author = "Feng, Jonathan L. and Tu, Huitzu and Yu, Hai-Bo",
    title = "{Thermal Relics in Hidden Sectors}",
    eprint = "0808.2318",
    archivePrefix = "arXiv",
    primaryClass = "hep-ph",
    reportNumber = "UCI-TR-2008-26",
    doi = "10.1088/1475-7516/2008/10/043",
    journal = "JCAP",
    volume = "10",
    pages = "043",
    year = "2008"
}

@article{Belanger:2021lwd,
    author = "Belanger, G. and Mjallal, A. and Pukhov, A.",
    title = "{Two dark matter candidates: The case of inert doublet and singlet scalars}",
    eprint = "2108.08061",
    archivePrefix = "arXiv",
    primaryClass = "hep-ph",
    doi = "10.1103/PhysRevD.105.035018",
    journal = "Phys. Rev. D",
    volume = "105",
    number = "3",
    pages = "035018",
    year = "2022"
}

@article{Belanger:2022esk,
    author = "B{\'e}langer, Genevi{\`e}ve and Pukhov, Alexander and Yaguna, Carlos E. and Zapata, {\'O}scar",
    title = "{The Z$_{7}$ model of three-component scalar dark matter}",
    eprint = "2212.07488",
    archivePrefix = "arXiv",
    primaryClass = "hep-ph",
    doi = "10.1007/JHEP03(2023)100",
    journal = "JHEP",
    volume = "03",
    pages = "100",
    year = "2023"
}

@article{Navarro:1996gj,
    author = "Navarro, Julio F. and Frenk, Carlos S. and White, Simon D. M.",
    title = "{A Universal density profile from hierarchical clustering}",
    eprint = "astro-ph/9611107",
    archivePrefix = "arXiv",
    doi = "10.1086/304888",
    journal = "Astrophys. J.",
    volume = "490",
    pages = "493--508",
    year = "1997"
}

@article{Bharucha:2022lty,
    author = {Bharucha, A. and Br{\"u}mmer, F. and Desai, N. and Mutzel, S.},
    title = "{Axion-like particles as mediators for dark matter: beyond freeze-out}",
    eprint = "2209.03932",
    archivePrefix = "arXiv",
    primaryClass = "hep-ph",
    doi = "10.1007/JHEP02(2023)141",
    journal = "JHEP",
    volume = "02",
    pages = "141",
    year = "2023"
}

@article{Slatyer:2015jla,
    author = "Slatyer, Tracy R.",
    title = "{Indirect dark matter signatures in the cosmic dark ages. I. Generalizing the bound on s-wave dark matter annihilation from Planck results}",
    eprint = "1506.03811",
    archivePrefix = "arXiv",
    primaryClass = "hep-ph",
    reportNumber = "MIT-CTP-4682",
    doi = "10.1103/PhysRevD.93.023527",
    journal = "Phys. Rev. D",
    volume = "93",
    number = "2",
    pages = "023527",
    year = "2016"
}

@article{Planck:2018vyg,
    author = "Aghanim, N. and others",
    collaboration = "Planck",
    title = "{Planck 2018 results. VI. Cosmological parameters}",
    eprint = "1807.06209",
    archivePrefix = "arXiv",
    primaryClass = "astro-ph.CO",
    doi = "10.1051/0004-6361/201833910",
    journal = "Astron. Astrophys.",
    volume = "641",
    pages = "A6",
    year = "2020",
    note = "[Erratum: Astron.Astrophys. 652, C4 (2021)]"
}

@article{Fermi-LAT:2015att,
    author = "Ackermann, M. and others",
    collaboration = "Fermi-LAT",
    title = "{Searching for Dark Matter Annihilation from Milky Way Dwarf Spheroidal Galaxies with Six Years of Fermi Large Area Telescope Data}",
    eprint = "1503.02641",
    archivePrefix = "arXiv",
    primaryClass = "astro-ph.HE",
    reportNumber = "FERMILAB-PUB-15-081-AE",
    doi = "10.1103/PhysRevLett.115.231301",
    journal = "Phys. Rev. Lett.",
    volume = "115",
    number = "23",
    pages = "231301",
    year = "2015"
}

@article{LZ:2024zvo,
    author = "Aalbers, J. and others",
    collaboration = "LZ",
    title = "{Dark Matter Search Results from 4.2 Tonne-Years of Exposure of the LUX-ZEPLIN (LZ) Experiment}",
    eprint = "2410.17036",
    archivePrefix = "arXiv",
    primaryClass = "hep-ex",
    reportNumber = "FERMILAB-PUB-24-0796-V",
    month = "10",
    year = "2024"
}

@article{PandaX-4T:2021bab,
    author = "Meng, Yue and others",
    collaboration = "PandaX-4T",
    title = "{Dark Matter Search Results from the PandaX-4T Commissioning Run}",
    eprint = "2107.13438",
    archivePrefix = "arXiv",
    primaryClass = "hep-ex",
    doi = "10.1103/PhysRevLett.127.261802",
    journal = "Phys. Rev. Lett.",
    volume = "127",
    number = "26",
    pages = "261802",
    year = "2021"
}

@article{Knowles:2021nvz,
    author = "Knowles, K. and others",
    title = "{The MeerKAT Galaxy Cluster Legacy Survey - I. Survey Overview and Highlights}",
    eprint = "2111.05673",
    archivePrefix = "arXiv",
    primaryClass = "astro-ph.GA",
    doi = "10.1051/0004-6361/202141488",
    journal = "Astron. Astrophys.",
    volume = "657",
    pages = "A56",
    year = "2022"
}

@inproceedings{Beck:2023oza,
    author = "Beck, Geoff and Makhathini, Sphesihle",
    title = "{Just a MeerKAT, or a dark matter machine?}",
    eprint = "2301.07910",
    archivePrefix = "arXiv",
    primaryClass = "hep-ph",
    month = "1",
    year = "2023"
}

@article{Djouadi:2005gi,
    author = "Djouadi, Abdelhak",
    title = "{The Anatomy of electro-weak symmetry breaking. I: The Higgs boson in the standard model}",
    eprint = "hep-ph/0503172",
    archivePrefix = "arXiv",
    reportNumber = "LPT-ORSAY-05-17",
    doi = "10.1016/j.physrep.2007.10.004",
    journal = "Phys. Rept.",
    volume = "457",
    pages = "1--216",
    year = "2008"
}

@article{Djouadi:2005gj,
    author = "Djouadi, Abdelhak",
    title = "{The Anatomy of electro-weak symmetry breaking. II. The Higgs bosons in the minimal supersymmetric model}",
    eprint = "hep-ph/0503173",
    archivePrefix = "arXiv",
    reportNumber = "LPT-ORSAY-05-18",
    doi = "10.1016/j.physrep.2007.10.005",
    journal = "Phys. Rept.",
    volume = "459",
    pages = "1--241",
    year = "2008"
}

@article{Gondolo:1990dk,
    author = "Gondolo, Paolo and Gelmini, Graciela",
    title = "{Cosmic abundances of stable particles: Improved analysis}",
    reportNumber = "UCLA-90-TEP-68",
    doi = "10.1016/0550-3213(91)90438-4",
    journal = "Nucl. Phys. B",
    volume = "360",
    pages = "145--179",
    year = "1991"
}

@article{Saikawa:2018rcs,
    author = "Saikawa, Ken'ichi and Shirai, Satoshi",
    title = "{Primordial gravitational waves, precisely: The role of thermodynamics in the Standard Model}",
    eprint = "1803.01038",
    archivePrefix = "arXiv",
    primaryClass = "hep-ph",
    reportNumber = "IPMU18-0037, MPP-2018-19",
    doi = "10.1088/1475-7516/2018/05/035",
    journal = "JCAP",
    volume = "05",
    pages = "035",
    year = "2018"
}

@article{Binder:2022pmf,
    author = "Binder, Tobias and Chakraborti, Sreemanti and Matsumoto, Shigeki and Watanabe, Yu",
    title = "{A global analysis of resonance-enhanced light scalar dark matter}",
    eprint = "2205.10149",
    archivePrefix = "arXiv",
    primaryClass = "hep-ph",
    reportNumber = "LAPTH-029/22",
    doi = "10.1007/JHEP01(2023)106",
    journal = "JHEP",
    volume = "01",
    pages = "106",
    year = "2023"
}

@article{Profumo:2017obk,
    author = "Profumo, Stefano and Queiroz, Farinaldo S. and Silk, Joseph and Siqueira, Clarissa",
    title = "{Searching for Secluded Dark Matter with H.E.S.S., Fermi-LAT, and Planck}",
    eprint = "1711.03133",
    archivePrefix = "arXiv",
    primaryClass = "hep-ph",
    doi = "10.1088/1475-7516/2018/03/010",
    journal = "JCAP",
    volume = "03",
    pages = "010",
    year = "2018"
}

@article{Su:2025mxv,
    author = "Su, Yu-Hang and Cai, Chengfeng and Zhang, Hong-Hao",
    title = "{Constraining secluded and catalyzed-annihilation dark matter models with Fermi-LAT and Planck data}",
    eprint = "2501.09647",
    archivePrefix = "arXiv",
    primaryClass = "hep-ph",
    doi = "10.1103/PhysRevD.111.075013",
    journal = "Phys. Rev. D",
    volume = "111",
    number = "7",
    pages = "075013",
    year = "2025"
}

@article{Datta:2023ncp,
    author = "Datta, AseshKrishna and Roy, Sourov and Saha, Abhijit Kumar and Tapadar, Ananya",
    title = "{An EFT origin of Secluded Dark Matter}",
    eprint = "2312.17171",
    archivePrefix = "arXiv",
    primaryClass = "hep-ph",
    month = "12",
    year = "2023"
}

@article{Du:2020avz,
    author = "Du, Yong and Huang, Fei and Li, Hao-Lin and Yu, Jiang-Hao",
    title = "{Freeze-in Dark Matter from Secret Neutrino Interactions}",
    eprint = "2005.01717",
    archivePrefix = "arXiv",
    primaryClass = "hep-ph",
    reportNumber = "ACFI-T20-04, UCI-EP-TR-2020-09",
    doi = "10.1007/JHEP12(2020)207",
    journal = "JHEP",
    volume = "12",
    pages = "207",
    year = "2020"
}

@article{ANTARES:2022aoa,
    author = "Albert, A. and others",
    collaboration = "ANTARES",
    title = "{Search for secluded dark matter towards the Galactic Centre with the ANTARES neutrino telescope}",
    eprint = "2203.06029",
    archivePrefix = "arXiv",
    primaryClass = "astro-ph.HE",
    doi = "10.1088/1475-7516/2022/06/028",
    journal = "JCAP",
    volume = "06",
    number = "06",
    pages = "028",
    year = "2022"
}

@article{Natarajan:2015hma,
    author = "Natarajan, Aravind and Aguirre, James E. and Spekkens, Kristine and Mason, Brian S.",
    title = "{Green Bank Telescope Constraints on Dark Matter Annihilation in Segue I}",
    eprint = "1507.03589",
    archivePrefix = "arXiv",
    primaryClass = "astro-ph.CO",
    month = "7",
    year = "2015"
}

@article{Booth:2009ex,
    author = "Booth, R. S. and de Blok, W. J. G. and Jonas, J. L. and Fanaroff, B.",
    title = "{MeerKAT Key Project Science, Specifications, and Proposals}",
    eprint = "0910.2935",
    archivePrefix = "arXiv",
    primaryClass = "astro-ph.IM",
    month = "10",
    year = "2009"
}

@article{Cembranos:2019noa,
    author = "Cembranos, J. A. R. and De La Cruz-Dombriz, {\'A} and Gammaldi, V. and M{\'e}ndez-Isla, M.",
    title = "{SKA-Phase 1 sensitivity to synchrotron radio emission from multi-TeV Dark Matter candidates}",
    eprint = "1905.11154",
    archivePrefix = "arXiv",
    primaryClass = "hep-ph",
    doi = "10.1016/j.dark.2019.100448",
    journal = "Phys. Dark Univ.",
    volume = "27",
    pages = "100448",
    year = "2020"
}

@article{Lavis:2023jju,
    author = "Lavis, Natasha and Sarkis, Michael and Beck, Geoff and Knowles, Kenda",
    title = "{Radio-frequency WIMP search with the MeerKAT galaxy cluster legacy survey}",
    eprint = "2308.08351",
    archivePrefix = "arXiv",
    primaryClass = "astro-ph.CO",
    doi = "10.1103/PhysRevD.108.123536",
    journal = "Phys. Rev. D",
    volume = "108",
    number = "12",
    pages = "123536",
    year = "2023"
}

@article{Sarkis:2024zjg,
    author = "Sarkis, Michael and Beck, Geoff",
    title = "{DarkMatters: A powerful tool for WIMPy analysis}",
    eprint = "2408.07053",
    archivePrefix = "arXiv",
    primaryClass = "hep-ph",
    doi = "10.1016/j.dark.2024.101745",
    journal = "Phys. Dark Univ.",
    volume = "47",
    pages = "101745",
    year = "2025"
}

@article{Kawasaki:2017bqm,
    author = "Kawasaki, Masahiro and Kohri, Kazunori and Moroi, Takeo and Takaesu, Yoshitaro",
    title = "{Revisiting Big-Bang Nucleosynthesis Constraints on Long-Lived Decaying Particles}",
    eprint = "1709.01211",
    archivePrefix = "arXiv",
    primaryClass = "hep-ph",
    reportNumber = "KEK-COSMO-211, IPMU17-0117, UT-17-29, KEK-Cosmo-211, KEK-TH-1998",
    doi = "10.1103/PhysRevD.97.023502",
    journal = "Phys. Rev. D",
    volume = "97",
    number = "2",
    pages = "023502",
    year = "2018"
}

@article{Kawasaki:2020qxm,
    author = "Kawasaki, Masahiro and Kohri, Kazunori and Moroi, Takeo and Murai, Kai and Murayama, Hitoshi",
    title = "{Big-bang nucleosynthesis with sub-GeV massive decaying particles}",
    eprint = "2006.14803",
    archivePrefix = "arXiv",
    primaryClass = "hep-ph",
    reportNumber = "KEK-Cosmo-254, KEK-TH-2214",
    doi = "10.1088/1475-7516/2020/12/048",
    journal = "JCAP",
    volume = "12",
    pages = "048",
    year = "2020"
}

@article{Coffey:2020oir,
    author = "Coffey, John and Forestell, Lindsay and Morrissey, David E. and White, Graham",
    title = "{Cosmological Bounds on sub-GeV Dark Vector Bosons from Electromagnetic Energy Injection}",
    eprint = "2003.02273",
    archivePrefix = "arXiv",
    primaryClass = "hep-ph",
    doi = "10.1007/JHEP07(2020)179",
    journal = "JHEP",
    volume = "07",
    pages = "179",
    year = "2020"
}

@article{1110.2895,
    author = "Cadamuro, Davide and Redondo, Javier",
    title = "{Cosmological bounds on pseudo Nambu-Goldstone bosons}",
    eprint = "1110.2895",
    archivePrefix = "arXiv",
    primaryClass = "hep-ph",
    reportNumber = "MPP-2011-116",
    doi = "10.1088/1475-7516/2012/02/032",
    journal = "JCAP",
    volume = "02",
    pages = "032",
    year = "2012"
}

@article{Nomura:2008ru,
    author = "Nomura, Yasunori and Thaler, Jesse",
    title = "{Dark Matter through the Axion Portal}",
    eprint = "0810.5397",
    archivePrefix = "arXiv",
    primaryClass = "hep-ph",
    doi = "10.1103/PhysRevD.79.075008",
    journal = "Phys. Rev. D",
    volume = "79",
    pages = "075008",
    year = "2009"
}

@article{Protheroe:1994dt,
    author = "Protheroe, R. J. and Stanev, T. and Berezinsky, V. S.",
    title = "{Electromagnetic cascades and cascade nucleosynthesis in the early universe}",
    eprint = "astro-ph/9409004",
    archivePrefix = "arXiv",
    reportNumber = "ADP-AT-94-5, LNGS-94-106",
    doi = "10.1103/PhysRevD.51.4134",
    journal = "Phys. Rev. D",
    volume = "51",
    pages = "4134--4144",
    year = "1995"
}

@article{Kawasaki:1994sc,
    author = "Kawasaki, M. and Moroi, T.",
    title = "{Electromagnetic cascade in the early universe and its application to the big bang nucleosynthesis}",
    eprint = "astro-ph/9412055",
    archivePrefix = "arXiv",
    reportNumber = "TU-474, ICRR-333-94-28",
    doi = "10.1086/176324",
    journal = "Astrophys. J.",
    volume = "452",
    pages = "506",
    year = "1995"
}

@article{Chakraborti:2018lso,
    author = "Chakraborti, Sreemanti and Poulose, Poulose",
    title = "{Interplay of Scalar and Fermionic Components in a Multi-component Dark Matter Scenario}",
    eprint = "1808.01979",
    archivePrefix = "arXiv",
    primaryClass = "hep-ph",
    doi = "10.1140/epjc/s10052-019-6933-y",
    journal = "Eur. Phys. J. C",
    volume = "79",
    number = "5",
    pages = "420",
    year = "2019"
}

@article{Chakraborti:2018aae,
    author = "Chakraborti, Sreemanti and Dutta Banik, Amit and Islam, Rashidul",
    title = "{Probing Multicomponent Extension of Inert Doublet Model with a Vector Dark Matter}",
    eprint = "1810.05595",
    archivePrefix = "arXiv",
    primaryClass = "hep-ph",
    doi = "10.1140/epjc/s10052-019-7165-x",
    journal = "Eur. Phys. J. C",
    volume = "79",
    number = "8",
    pages = "662",
    year = "2019"
}

@article{Wang:2023sxr,
    author = "Wang, Guan-Sen and Chen, Zhan-Fang and Zu, Lei and Gong, Hao and Feng, Lei and Fan, Yi-Zhong",
    title = "{SKA sensitivity for possible radio emission from dark matter in Omega Centauri}",
    eprint = "2303.14117",
    archivePrefix = "arXiv",
    primaryClass = "astro-ph.HE",
    doi = "10.1088/1475-7516/2024/05/129",
    journal = "JCAP",
    volume = "05",
    pages = "129",
    year = "2024"
}

@article{Chatterjee:2025vdz,
    author = "Chatterjee, Shiuli and Hryczuk, Andrzej",
    title = "{Conversions in two-component dark sectors: a phase space level analysis}",
    eprint = "2502.08725",
    archivePrefix = "arXiv",
    primaryClass = "hep-ph",
    doi = "10.1007/JHEP07(2025)279",
    journal = "JHEP",
    volume = "07",
    pages = "279",
    year = "2025"
}

@article{Borah:2025fkd,
    author = "Borah, Debasish and Mahapatra, Satyabrata and Nanda, Dibyendu and Sahoo, Sujit Kumar and Sahu, Narendra",
    title = "{Effective theory of light Dirac neutrino portal dark matter with observable {\ensuremath{\Delta}}Neff}",
    eprint = "2502.10318",
    archivePrefix = "arXiv",
    primaryClass = "hep-ph",
    doi = "10.1103/m7my-1cjy",
    journal = "Phys. Rev. D",
    volume = "112",
    number = "5",
    pages = "055010",
    year = "2025"
}

@article{Visinelli:2015eka,
    author = "Visinelli, Luca and Gondolo, Paolo",
    title = "{Kinetic decoupling of WIMPs: analytic expressions}",
    eprint = "1501.02233",
    archivePrefix = "arXiv",
    primaryClass = "astro-ph.CO",
    doi = "10.1103/PhysRevD.91.083526",
    journal = "Phys. Rev. D",
    volume = "91",
    number = "8",
    pages = "083526",
    year = "2015"
}

@article{Evans:2019vxr,
    author = "Evans, Jared A. and Gaidau, Cristian and Shelton, Jessie",
    title = "{Leak-in Dark Matter}",
    eprint = "1909.04671",
    archivePrefix = "arXiv",
    primaryClass = "hep-ph",
    doi = "10.1007/JHEP01(2020)032",
    journal = "JHEP",
    volume = "01",
    pages = "032",
    year = "2020"
}

@article{Fernandez:2021iti,
    author = "Fernandez, Nicolas and Kahn, Yonatan and Shelton, Jessie",
    title = "{Freeze-in, glaciation, and UV sensitivity from light mediators}",
    eprint = "2111.13709",
    archivePrefix = "arXiv",
    primaryClass = "hep-ph",
    doi = "10.1007/JHEP07(2022)044",
    journal = "JHEP",
    volume = "07",
    pages = "044",
    year = "2022"
}

@article{Berger:2016vxi,
    author = "Berger, Joshua and Jedamzik, Karsten and Walker, Devin G. E.",
    title = "{Cosmological Constraints on Decoupled Dark Photons and Dark Higgs}",
    eprint = "1605.07195",
    archivePrefix = "arXiv",
    primaryClass = "hep-ph",
    reportNumber = "SLAC-PUB-16533",
    doi = "10.1088/1475-7516/2016/11/032",
    journal = "JCAP",
    volume = "11",
    pages = "032",
    year = "2016"
}
\bibliographystyle{JHEP}
\end{document}